\documentclass[journal]{IEEEtran}
\usepackage{amsmath,amsfonts}
\usepackage{algorithmic}
\usepackage{algorithm}
\usepackage{array}
\usepackage[caption=false,font=normalsize,labelfont=sf,textfont=sf]{subfig}
\usepackage{textcomp}
\usepackage{stfloats}
\usepackage{url}
\usepackage{verbatim}
\usepackage{graphicx}
\usepackage{cite}
\usepackage{multirow}
\usepackage{color}
\usepackage{hyperref}
\newcommand*{\rom}[1]{\uppercase\expandafter{\romannumeral#1\relax}}
\begin{document}

\title{SNR-Independent Joint Source-Channel Coding for wireless image transmission }

\author{Hongjie Yuan, Weizhang Xu, Yuhuan Wang, Xingxing Wang

    \thanks{The authors are with the State Key Laboratory of Media Convergence \& Communication, Communication University of China, Beijing 100024, China. Weizhang Xu is the Corresponding author (e-mail: wzhxu@cuc.edu.cn)}
    \thanks{This work is supported by “the Fundamental Research Funds for the Central Universities”}

}



\IEEEoverridecommandlockouts
\IEEEpubidadjcol

\IEEEpubid{\begin{minipage}{\textwidth}\ \\[50pt] \centering
\copyright This work has been submitted to the IEEE for possible publication. Copyright may be transferred without notice, after which this version may no longer be accessible.
\end{minipage}} 

\maketitle
\IEEEpubidadjcol

\begin{abstract}
    Significant progress has been made in wireless Joint Source-Channel Coding (JSCC) using deep learning techniques.
    The latest DL-based image JSCC methods have demonstrated exceptional performance during transmission, while also avoiding cliff effects.
    However, current channel adaptive JSCC methods rely on channel SNR information, which can lead to performance degradation in practical applications due to channel mismatch effects.
    This paper proposes a novel approach for image transmission, called SNR Independent Joint Source-Channel Coding (SIJSCC), which utilizes Deep Learning techniques to achieve exceptional performance across various signal-to-noise ratio (SNR) levels without SNR estimating.
    We have designed an Inverted Residual Attention Bottleneck (IRAB) module for the model, which can effectively reduce the number of parameters while expanding the receptive field.
    In addition, we have incorporated a convolution and self-attention mixed encoding module to establish long-range dependency relationships between channel symbols.
    Our experiments have shown that SIJSCC outperforms existing channel adaptive DL-based JSCC methods that rely on SNR information.
    Furthermore, we found that SNR estimation does not significantly benefit SIJSCC, which provides insights for the future design of DL-based JSCC methods.
    The reliability of the proposed method is further demonstrated through an analysis of the model bottleneck and its adaptability to different domains, as shown by our experiments.
\end{abstract}


\section{Introduction}
\begin{figure}[!t]
    \centering
    \includegraphics[width=\linewidth]{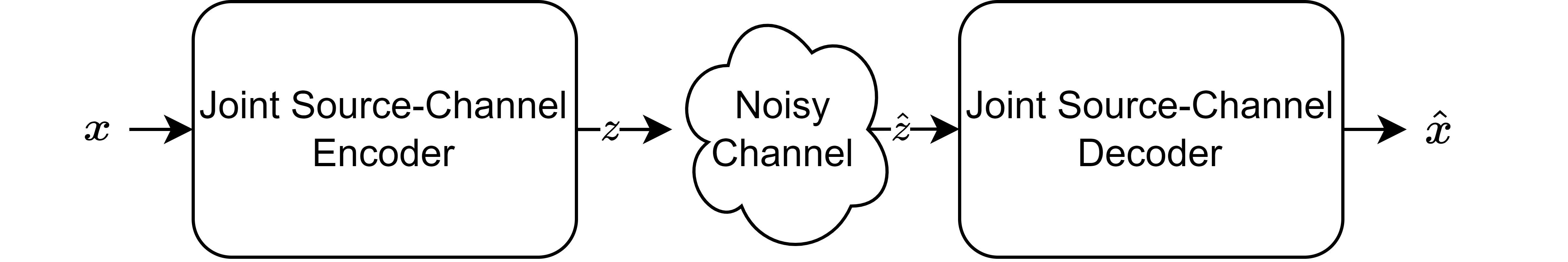}
    \caption{ Block diagram of the DL-based JSCC pipeline.}
    \label{fig:dl-jscc}
\end{figure}

\IEEEPARstart{D}{espite} being theoretically proven to have superior performance, JSCC has a challenging balance between design complexity and performance, and thus is not widely adapted in practical applications at present \cite{gastpar_code_2003, kostina_joint_2017, skoglund_design_2002,mittal_hybrid_2002, chatellier_robust_2007,kozintsev_robust_1998 }.
With the advancement of deep learning technology, the data-driven DL-based JSCC has recently attracted great interest from researchers\cite{yang_introduction_2022,xu_wireless_2022, yang_deep_2021, wang_novel_2021,burth_kurka_joint_2020,bourtsoulatze_deep_2019,kurka_deepjscc-f_2020, yang_deep_2022,sun_joint_2022 ,ding_snr-adaptive_2021}. The intelligent and concise design concept of DL-based JSCC aligns with the fundamental purpose of communication.
Fig. \ref{fig:dl-jscc} depicts the recently proposed DL-based image JSCC algorithm \cite{bourtsoulatze_deep_2019}.
As in computer vision, where Deep Neural Networks has replaced handcrafted feature extraction in \cite{simonyan_very_2015}, DL-based JSCC has replace the probability model in traditional JSCC tasks with an implicit model. These DL-based methods discard explicit source and channel encoding schemes and instead employ a neural network to map pixel values to the channel input signal. The performance of DL-based JSCC is directly affected by the presence of channel noise, resulting in a gradual decrease in image reconstruction quality as the SNR of the channel decreases.

Recently, several remarkable studies have emerged in the field of DL-based JSCC. Some researchers have developed DL-based JSCC schemes that incorporate traditional channel coding principles. Kurka et al. \cite{kurka_deepjscc-f_2020}  drew inspiration from the Schalkwijk-Kailath scheme, they proposed Autoencoder(AE) based JSCC scheme employs a multi-level feedback. This approach involves dividing the transmission of each image into L layers, with each layer aiming to enhance the accuracy of the receiver's estimation by transmitting additional information pertaining to residual errors.
Inspired by the turbo principle, \cite{wang_novel_2021} presents an iterative receiver for reconstructing images using turbo codes. They use a refinement neural network to adjust the output of the Neural Source Decode (NSD) and use EXIT charts to predict message reliability.
Schalkwijk et al.\cite{yang2022resolution} were inspired by the source coding principle and proposed a Resolution-adaptive Deep JSCC(RaDJSCC). This approach integrates DL-based JSCC architecture with the quadtree structure region rate allocation strategy used in the HEVC standard \cite{liu_learned_2021}. By incorporating this integration, RaDJSCC is able to perceive the content of the transmitted image and adaptively allocate more channel bandwidth to complex pixel blocks.
In addition to providing higher image transmission quality, DL-based JSCC also exhibits good scalability. For instance, Yang et al.\cite{yang_deep_2022} proposed a dynamic rate control scheme for deep encoding and decoding that uses SNR feedback and the Gumbel-Softmax trick to select the transmitted feature layer.

The problem with the aforementioned DL-based JSCC approach is that the model needs to match the current communication channel strictly. In practical deployment, frequent switching of the model due to channel variations may lead to increased system overhead and reduced accuracy.
To tackle this problem, Bao et al.\cite{bao_adjscc-l_2021} proposed Attention DL based JSCC (ADJSCC), which enables a single model to adapt to various channel conditions. This design has also been adapted in \cite{yang_deep_2022} and \cite{wu_channel-adaptive_2022}.
Compared to models that adapt to the channel independently, ADJSCC demonstrates better robustness.
However,  this approach still has some challenges in practical applications.
The channel feedback-based scheme may experience significant performance degradation in reconstruction performance due to the channel mismatch effect\cite{kim_deepcode_2020}.
Additionally, the channel feedback-based scheme is not suitable for wireless broadcast transmission. Ding et al. proposed a channel adaptive JSCC scheme without the channel feedback module, which can be applied to multi-user scenarios.
However, the performance of this scheme is not as effective as the channel feedback-based scheme, and it still relies on SNR estimation.
Therefore, in most cases, it is not possible to achieve the same performance as in simulation experiments with perfect channel feedback.

We have reviewed the state-of-the-art DL-based JSCC schemes and the role of channel state information. Firstly, it is difficult to balance the benefits of channel estimation and the drawbacks of channel mismatch in DL-based JSCC. Secondly, channel feedback-based schemes are not suitable for wireless broadcast transmission.
Shannon has pointed out that the feedback of SNR cannot increase the channel capacity of the network \cite{shannon_zero_1956}.
Our objective is to develop a DL-based JSCC method that can transmit data in a dynamic channel without relying on a SNR information.
This presents a challenging problem, as it requires the encoder to compress data while introducing redundancy to account for channel uncertainty in scenarios where the channel capacity is unknown. Additionally, the decoder needs to recover the original signal from the received data in the presence of unknown SNR. Accurate estimation and inference of channel characteristics are essential for achieving efficient data transmission and reconstruction.

This paper proposes a DL-based JSCC model, referred to as the SNR Independent JSCC (SIJSCC) model.
To address the challenge of determining the appropriate encoding and decoding strategy without SNR of the channel, SIJSCC employs carefully designed encoder and decoder architectures.
We proposed the Inverted Residual Attention Bottleneck (IRAB) as its fundamental structure. The IRAB structure is designed to be lightweight, reducing computational and memory requirements.
Additionally, by incorporating an enhanced spatial attention(ESA)\cite{woo_convnext_2023,liu_residual_2020,kong_residual_2022} module, IRAB can effectively decouple information on local spatial features and allocate resources to more significant regions.
In addition to this, SIJSCC has integrated a self-attention module \cite{pan_integration_2022} into both the encoder output and decoder input of the proposed SIJSCC architecture, facilitating the establishment of long-range correlations among the encoded symbols. This integration is particularly meaningful, especially when considering the varying information density and similarity among different regions within an image.
The experimental results show that SIJSCC can achieve better performance than the state-of-the-art DL-based JSCC schemes without channel information. We have also discovered that SIJSCC is capable of effectively learning channel characteristics while maintaining robustness.
Our main contributions can be summarized as follows:
\begin{itemize}
    \item We propose a novel DL-based JSCC encoder structure, SIJSCC, which enables image transmission over varying wireless channels while mitigating the effects of channel mismatch.
    \item SIJSCC follows Autoencoder (AE) principles, which results in a simple overall structure. This simplicity makes SIJSCC suitable for broadcast communication, and facilitates its practical deployment in various applications.
    \item SIJSCC encompasses meticulously designed network structures. We proposed an IRAB module as the fundamental structure of SIJSCC and integrated a self-attention module. This design empowers SIJSCC with robust encoding and decoding capabilities while minimizing resource consumption, thereby surpassing the current state-of-the-art DL-based JSCC methods comprehensively.
    \item  The effectiveness of SIJSCC has been demonstrated through numerical experiments. Specifically, we have validated its domain adaptation capability, as training datasets often fail to align with real-world data. Experimental results show that SIJSCC demonstrates exceptional domain adaptation capabilities.
\end{itemize}

The rest of this work is organized as follows. In section \ref{sec:related}, we review the related research on SIJSCC. Section \ref{sec:method} introduces the proposed SIJSCC model. Section \ref{sec:experiments} presents the experimental results and analysis. Finally, we conclude this paper in section \ref{sec:conclusion}.

\section{Related Works}\label{sec:related}
\subsection{Autoencoder Framework}

The General Autoencoder Framework proposed by Hinton et al. has been used to address the issue of lacking supervision in backpropagation \cite{rumelhart_learning_1988, hinton_reducing_2006}. The corresponding AE problem aims to find a compressed representation of the input data while minimizing the reconstruction error between the original input and the reconstructed output. Autoencoder has become an important component in deep learning\cite{baldi_autoencoders_2012, balle_end--end_2017}, and can be utilized for data compression and feature extraction in JSCC.

Given an input space and a feature space, AE tries to find the mappings between them such that the reconstruction error of the input features is minimized.
Let $\mathbb{X}$ and $\mathbb{Z}$ denote two sets of collections respectively, $\mathcal{F}$ denotes to a class of functions that map elements in $\mathbb{X}$ to the latent space $\mathbb{Z}$, $\mathcal{G}$ represents a class of functions that remap elements in $\mathbb{Z}$ back to $\mathbb{X}$, and $\Delta$ is a distortion function that can be either discrete or continuous. For arbitrary $f\in\mathcal{F}$ and $g\in\mathcal{G}$, the autoencoder problem can be stated as follows:
\begin{equation}
    \begin{aligned}
         & \mathcal{F}: \mathbb{X} \rightarrow \mathbb{Z}                                                               \\
         & \mathcal{G}: \mathbb{Z} \rightarrow \mathbb{X}                                                               \\
         & \left(f^*, g^*\right)=\arg \min _{f \in \mathcal{F}, g \in \mathcal{G}} \Delta(\mathbb{X}, g[f(\mathbb{X})])
    \end{aligned}
\end{equation}
Here, we only consider the auto-associative case where the input and output targets are the same.

More closely related to JSCC is the Denoising Autoencoder(DAE) \cite{vincent_extracting_2008}, which is based on the assumption that a good representation is expected to capture stable structures, in the form of dependencies and regularities, that are present in the distribution of observed inputs. The objective optimization equation for DAE can be expressed as follows:
\begin{equation}
    \left(f^*, g^*\right)=\arg \min _{f \in \mathcal{F}, g \in \mathcal{G}} \Delta(\mathbb{X}, g[f(\widehat{\mathbb{X}})])
\end{equation}
The main difference between DAE and autoencoder is that DAE takes the input data from $\hat{x}\in\hat{\mathbb{X}}$, which is obtained by randomly corrupting the true data $x$ as $\hat{x}=x\odot q_D\left(\hat{x}\middle| x\right)$, allowing the model to learn better robust representation. The main difference between DAE and JSCC lies in the process of adding noise and the type of noise used. JSCC focuses more on the reversibility and reconstruction quality of the data, and the use of channel characteristics in compression processing, while DAE focuses more on the accuracy of reconstructing the original content rather than the reconstruct ability or stability of the signal.

\subsection{Attention Mechanism}

The attention mechanism is a key concept in deep learning, inspired by the human biological system. It rapidly scans global information and concentrates on target areas. These mechanisms include channel attention, spatial attention, or the independent reweighting of both channels and spatial locations to refine the feature map \cite{hu_squeeze-and-excitation_nodate,park_bam_2018,hu_gather-excite_2019,woo_cbam_2018,kim_mamnet_2018}. They improve the receptive field of convolutional networks by assigning weights to channels or spatial locations on the feature maps to focus more on significant local areas in the input data.
Given an intermediate feature map $W_{in} \in \mathbb{R}^{ C\times H \times W}$ as input,
the final output vector $W_{out}$:
\begin{equation}\label{eq:attention}
    W_{out} = M(W_{in})\bigotimes W_{in}
\end{equation}
where $\bigotimes$ denotes element-wise multiplication, and $A=M(W_{in})$ is the attention map.

The self-attention mechanism is a unique form of attention that was initially introduced in the Transformer model and primarily applied in natural language processing \cite{vaswani_attention_2017}. While convolutional attention is effective for capturing local dependencies, self-attention is a more flexible mechanism that can capture both local and global dependencies. The calculation process of self-attention mechanism can be represented as:
\begin{equation}
    \begin{aligned}
        \operatorname{Attention}(QW^Q,KW^K) & =\operatorname{softmax}\left(\frac{(QW^Q)(KW_K)^{T}}{\sqrt{d}}\right) \\
        W_{out}                             & =\operatorname{Attention}(QW_Q,KW_K)(VW_V)
    \end{aligned}
\end{equation}
Here, $Q$, $K$ and $V$ represent the Query, Key, and Value vector sequences obtained by linear transformation of the input feature map $W$, and $d$ represents the dimension of each vector. The softmax function is applied to the dot product of the query and key vectors, and the resulting attention weights are used to compute a weighted sum of the value vectors.

Multi-Head Self-Attention is an extended form of attention mechanism that introduces multiple attention heads in self-attention models to capture different points of focus. The calculation process of Multi-Head Self-Attention can be represented as:
\begin{equation}\label{eq:multi-head}
    \text{MultiHead}(Q, K, V) = \text{Concat}(\text{head}^1, \text{head}^2, ..., \text{head}^h)W_O
\end{equation}
where $W^Q_i$, $W^K_i$, and $W^V_i$ represent the projection matrices for the i-th of $h$ attention head:
\begin{equation}
    \text{head}^i = \operatorname{Attention}(QW_Q^i,KW_K^i)VW_V^i
\end{equation}
The introduction of multi-head self-attention enhances the expressive power of the model by allowing it to attend to different information through different attention heads. This enables the model to extract more comprehensive and rich features.

Although the self-attention mechanism has shown great performance in computer vision tasks, it also increases the training and inference time, and also requires higher computational resources\cite{dosovitskiy_image_2021, wang_non-local_2018}.  Recently, some research has achieved significant improvements in computer vision tasks by combining self-attention mechanism with convolutional networks, including reducing computational complexity, using convolutional operations to complement the Transformer model, and introducing additional inductive biases \cite{liu_swin_2021,bello_attention_2019,srinivas_bottleneck_2021}. The self-attention mechanism computes dot-product attention scores and performs weighted summation to obtain a more compact sequence representation that minimizes redundancy while retaining the semantic information of the original features. This makes the sequence representation easier to store, process, and transfer.

\section{Proposed method}\label{sec:method}

\subsection{Overall Architecture}

\begin{figure*}[!t]
    \centering
    \includegraphics[width=5.5in]{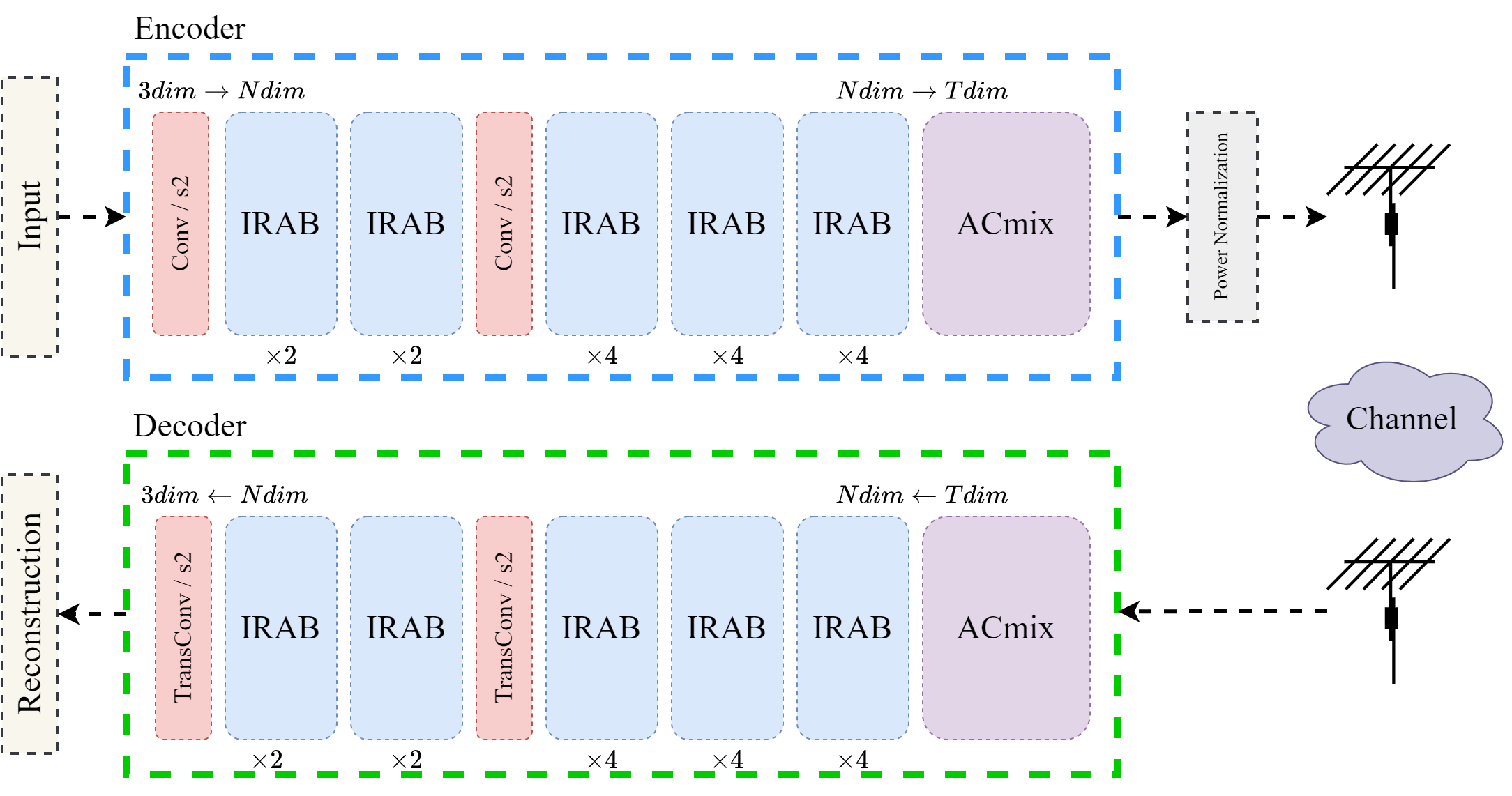}%

    \centering
    \caption{Network Architecture of the Proposed SIJSCC Method. The blue dashed box represents the encoder, while the green box represents the decoder. The numbers underneath each IRAB block indicate the bottleneck expansion factor.\ $N$ and $T$ represent the dimension of channels in the feature map.}
    \label{fig:SIJSCC}
\end{figure*}

We will now provide a detailed description of the design of SIJSCC. As illustrated in Fig. \ref{fig:SIJSCC}, SIJSCC consist of a single encoder and decoder, without any redundant branches.
The encoder directly maps the source data into discrete signals, which are then transmitted through an unknown wireless channel. The decoder maps the received discrete signals back to the source data. We believe that neural network-based encoding and decoding methods can autonomously learn the underlying patterns of the channel, thereby reducing the impact of channel noise and other errors.

Consider a wireless system for transmitting images $x$ that have a size of $h\left(height\right)\times w\left(width\right)\times c\left(channel\right)$, where $x\in\mathbb{R}^n$, and $n=w\times h\times c$. The encoder, parameterized by $\theta$, first maps input $x$ to $z$ through a deterministic mapping:
\begin{equation}
    z=f_\theta(x)\in\mathbb{C}^k
\end{equation}
During this process, we simply use two convolution layers with a stride of 2 to decrease the size of the feature map.
After passing through the first convolution layer of the encoder, the output feature dimension of the image is mapped to $N$. In this paper, the default value of $N$ is set to 192.
We utilize stacked IRAB blocks to extract features, where the input and output feature channel dimensions of the preceding IRAB blocks are equal to $N$. In the final IRAB block, the dimension of the feature map channels is changed to $T$.
At the output layer of the encoder, we use a self-attention module ACmix \cite{pan_integration_2022}, to capture the long distance relationship between transmitted signal. This design is inspired by Bottleneck Transformers \cite{srinivas_bottleneck_2021}, which effectively learn abstract and low-resolution feature maps from large images using convolution, then handle and aggregate the information contained in the feature maps captured by convolution layer using global self-attention.

The encoded image is then conjugate transposed to obtain the complex channel signal ${z}\in\mathbb{C}^k$, where $k$ represents the dimension of the encoded complex symbols.
We refer to the n as the source bandwidth and the k as the channel bandwidth, and the $Ratio=k/n$ is called the bandwidth compression rate.
SIJSCC adjusts the dimension $T$ of the final self attention module to control the transmission rates. As the encoder includes two down sampling operations with $stride=2$, the formula for $k$ is given by $k=T\times n/\left(2\times 16\times c\right)$. As the value of $c$ is typically 3, the bandwidth compression of SIJSCC is equal to $T/96$. To serve as the base line, in this paper we only consider the AWGN channel:
\begin{equation}
    \hat{z}={z}+\omega
\end{equation}
Where $\omega\in\mathbb{C}^k$ is a circularly symmetric complex Gaussian distribution noise.
The model can also be trained on other fading channel types, including Rayleigh and Rician fading channels, by using appropriately simulated channel conditions.
The received signal $\hat{z}$ is then send decoderthen mapped back to a “reconstructed” vector in input space. The encoder and decoder of SIJSCC are designed to be symmetric, and the decoder is the mirror image of the encoder. with the decoder parameterized by $\theta^{\prime}$:
\begin{equation}
    \hat{x}=g_{\theta^{\prime}}(\hat{z})\in\mathbb{R}^n
\end{equation}

SIJSCC is specifically designed for the transmission of images in dynamic wireless channels without the need for channel information. Thus $\omega$ is an independent variable that is unknown to the encoder and decoder. The entire process can be expressed as an optimization objective.

\begin{equation}
    \begin{aligned}
        \theta^{\star}, \theta^{\prime \star} & =\underset{\theta, \theta^{\prime}}{\arg \min } \frac{1}{n} \sum_{i=1}^n L\left({x}^{(i)}, \hat{x}^{(i)}\right)                                                          \\
                                              & =\underset{\theta, \theta^{\prime}}{\arg \min } \frac{1}{n} \sum_{i=1}^n L\left({x}^{(i)}, g_{\theta^{\prime}}\left(f_\theta\left({x}^{(i)}\right)+\omega \right)\right)
    \end{aligned}
\end{equation}
where $L$ is the loss function, and $\theta^{\star}$ and $\theta^{\prime \star}$ are the optimal parameters of the encoder and decoder, respectively.
SIJSCC can be considered as a specific case of an autoencoder. When the value of $\omega$ equal to $0$, SIJSCC is equivalent to a traditional autoencoder.
This design is easily deployable in practical environments. However, as mentioned in the introduction, it poses significant challenges to the encoding and decoding capabilities of the network.

\subsection{Inverted Residual Attention Bottleneck}
\begin{figure*}[!t]
    \centering
    \subfloat[]{\includegraphics[height=2.5in]{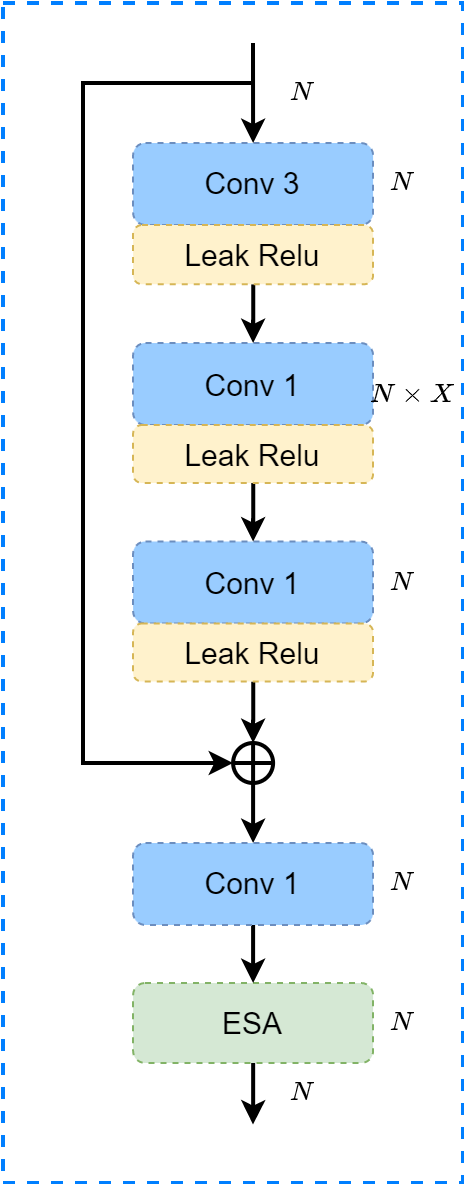}\label{fig:irab}}%
    \hfil
    \centering
    \subfloat[]{\includegraphics[height=2.5in]{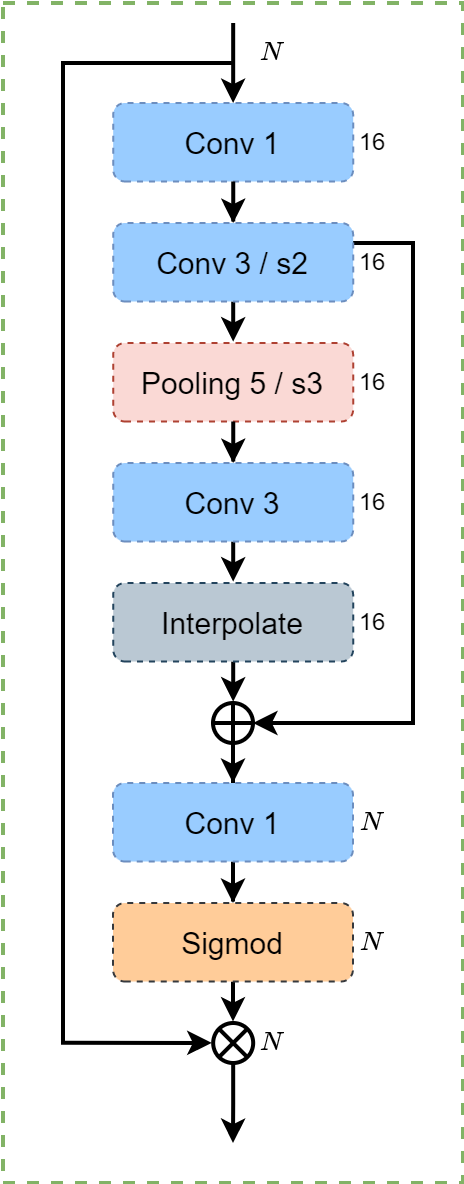}\label{fig:esa}}%
    \hfil
    \centering
    \subfloat[]{\includegraphics[height=3in]{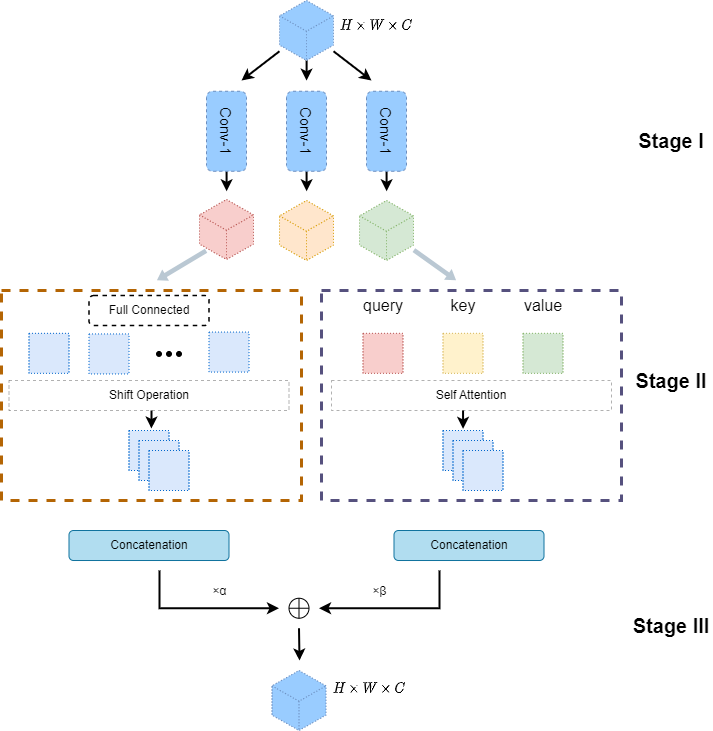}\label{fig:acmix}}%
    \caption{(a)Invented Residual Attention Bottleneck (IRAB). The numbers after "Conv" represent the current convolution kernel size. In IRAB, $X$ represents the multiple of the current convolution kernel size relative to the input. (b) Enhanced Spatial Attention block(ESA).  (c) Convolution and Attention Mixed (ACmix) block. It can be divided into three stages. In the second stage, the left box represents the convolution process, while the right box represents the self-attention process.}
    \label{fig:SIJSCC_modules}
\end{figure*}

As shown in Fig. \ref{fig:irab}, the IRAB consists of two stages. Firstly, it utilizes an inverted bottleneck residual module, inspired by the work of Liu et al. \cite{liu_convnet_2022}, to extract features. In this stage, the input tensor undergoes dense computation using a $3\times3$ convolution layer.
Then, an expanded $1\times1$ convolution layer is applied to increase the dimensions of the convolution layer by a factor of $X$. Finally, a $1\times1$ convolution layer is used to restore the tensor to its original dimensions.
When ignoring the bias term, the computational complexity of the inverted bottleneck residual design can be expressed as the ratio of $(2\times X+9)/27$ Flops to that of a regular $3\times3$ residual block.
In the first downsampling IRAB module, $X$ is set to 2, and in the second downsampling IRAB module, $X$ is set to 4. This design effectively reduces the computational burden on large-scale shallow information, improving computational efficiency.

Next, the tensor is passed through a $1\times1$ convolution and then input into an enhanced spatial attention (ESA) modules \cite{wang_residual_2017,liu_residual_2020}.
As illustrated in Figure \ref{fig:esa}, the ESA block initiates with a $1 \times 1$ convolutional layer to reduce the channel dimension from $N$ to 16, this helps in reducing the computational complexity. Subsequently, a convolution layer with a stride of 2, followed by a max-pooling layer, is employed to reduce the feature size. This downsampling of features aids in enlarging the receptive field of the convolution layers, facilitating a better allocation of attention weights. Afterwards, the spatial and channel dimensions of the input features are restored through interpolation and a $1 \times 1$ convolution layer. Finally, the attention weights of the input features are obtained using the sigmoid function $M$ as in Eq. \ref{eq:attention}. ESA follow two design principles. First, it is lightweight and can be embedded in each block. Second, it has a sufficiently large receptive field to ensure that attention mechanisms can fully learn high-frequency region information on the feature map.

The IRAB module only consists of highly optimized $1\times1$ and $3\times3$ convolution layers, which enable efficient feature extraction. Additionally, by incorporating an enhanced spatial attention module, it can effectively separate the mechanisms spatially and allocate more resources to regions with dense information.

\subsection{Convolution and Self-attention Mixed Module}

The inter-symbol correlations between the encoded symbols over long distances are often overlooked or weakened, this can potentially impact the accuracy and effectiveness of the encoding and decoding process.
Since ESA can only handle attention features within a local neighborhood.
Therefore, a self-attention module is used to capture encoded feature information over long distances. Here we adopted a modified attention and convolution hybrid structure ACmix \cite{pan_integration_2022}.
ACmix decouples the self-attention and convolution operations and has discovered that they share the same computation overhead when projecting the input feature maps.

As shown in Fig. \ref{fig:acmix}, ACmix can be divided to three stages. At the first stage, ACmix employs three $1\times1$ convolution layer to project the input feature maps into three feature maps.
The second stage corresponds to the process of aggregating features, where the intermediate features from the first stage are shared between the convolution and self-attention operations, despite their different paradigms.
For convolution path of ACmix, input feature maps is concentrated and projected into $k\times k$ feature maps, followed by shifting and aggregating the generated features. ACmix applying depthwise convolution with fixed kernels as a replacement of the inefficient tensor shifts.
For the self-attention path of ACmix, the input feature map is first projected into $N$ groups, with each group consisting queries, keys, and values obtained from the $1\times1$ convolutions. The attention weights are used to aggregate the values, following the conventional multi-head self-attention modules in Eq. \ref{eq:multi-head}.
Finally, in the last stage, the outputs of the two paths are added together, and the weights are controlled by two learnable parameters $\alpha$ and $\beta$.
This design ensures that traditional convolutions, which share convolution filter weights across the entire feature map, utilize the inherent features of the aggregation function in local receptive fields, thereby applying a critical inductive bias to image processing. Another advantage of using ACmix is that it can accept input feature maps of any size. In our experiments, we found that this structure performs better than pure self-attention network in JSCC tasks \cite{wang_non-local_2018}.

We replaced the batch normalization layer with a generalized divisive normalization(GDN) \cite{balle_density_2016}. It has been proven that GDN has a better fit to the pairwise statistics of local filter responses, resulting in more natural samples of image patches and better performance as a prior for image processing problems such as denoising.For more detailed information, please refer to the original paper.

\section{Experiments and Analysis }\label{sec:experiments}
\subsection{Training and Implementation Details}
The experiments employed the ImageNet ILSVRC 2017 as the training dataset \cite{russakovsky_imagenet_2015}. In each iteration, the images were randomly cropped into $128\times128$ patches. The model was trained on a single Nvidia RTX 4090, with a batch size of 112. The initial learning rate is set to 0.0001, and an exponentially decaying learning rate was employed based on the loss. Training was stopped when the model converged.

In previous studies, the mean square error (MSE) was widely used as a distortion function. However, Lim et al. \cite{lim_enhanced_2017} found that MSE has limitations in image reconstruction tasks. Therefore, in this paper, we chose the L1 Charbonnier loss function as the loss function. The L1 Charbonnier loss function is a smooth L1 loss function, also known as a smooth L1 norm loss function. Suppose  $x$ and $\hat{x}$  represent the original image and the reconstructed image, the definition of L1 Charbonnier is as follows:
\begin{equation}
    L\left(\hat{x},x\right)=\sqrt{\left(x-\hat{x}\right)^2+\epsilon}
\end{equation}
Where $\epsilon$ is a small value to ensure that the Charbonnier penalty is never zero. The loss function is smoother than normal L1 and therefore more robust to noise and outliers. In some image processing tasks involving noise, L1 Charbonnier loss function usually provides better results \cite{charbonnier_two_1994}, \cite{lai_deep_2017}.

During the training stage, we utilized the recently proposed Lion optimizer \cite{chen_symbolic_2023}.
Lion is an efficient optimization algorithm discovered using symbolic program search space and evolutionary search techniques. It only tracks momentum and uses the sign operation to compute updates, resulting in lower memory overhead compared to the widely used Adam optimizer\cite{kingma2014adam}.
Lion helps the model maintain relatively low error rates against adversarial Gaussian perturbations and reduces pre-training computational cost while enhancing training efficiency. We find that Lion optimizer can accelerate model convergence.

After training is completed, the paper utilizes Peak Signal-to-Noise Ratio (PSNR) and The Structural Similarity Index Measure (SSIM) as metrics to quantitatively measure the quality of image restoration. PSNR is a measure of the similarity between two images, and is calculated using the following formula:
\begin{equation}
    \mathrm{PSNR}(x,\hat x)=10\log_{10}{\left(\frac{{\mathrm{MAX}(x)}^2}{\mathrm{MSE}(x,\hat x)}\right)}
\end{equation}
Here, MAX is the maximum possible pixel value, for example, for an 8-bit image, MAX is 255. MSE is the mean squared error, and its calculation is defined by the following equation:
\begin{equation}
    \mathrm{MSE}(x, \hat x)=\frac{1}{mn}\sum_{i=0}^{m-1}\sum_{j=0}^{n-1}\left[x\left(i,j\right)-\hat{x}\left(i,j\right)\right]^2
\end{equation}
Where $m$ and $n$ represent the width and height of the images, respectively.
SSIM is another measure of the structural similarity between the original and processed images. It provides a more accurate measure of perceived image quality by taking into account structural information and human visual perception. The formula for calculating SSIM is as follows:
\begin{equation}
    \mathrm{SSIM}\left(x,\hat{x}\right)=\frac{\left(2\mu_x\mu_{\hat{x}}+c_1\right)\left(2\sigma_{x{\hat{x}}}+c_2\right)}{\left(\mu_x^2+\mu_{\hat{x}}^2+c_1\right)\left(\sigma_x^2+\sigma_{\hat{x}}^2+c_2\right)}
\end{equation}
Where $\mu$ is the mean of the image, $\sigma$ is the standard deviation of the image, $2\sigma_{x{\hat{x}}}$\ is the covariance of the image $x$ and ${\hat{x}}$, Additionally, $c_1$ and $c_2$ are constants used to avoid division by zero.

\subsection{Experimental and Results}
To the best of our knowledge, the most effective DL-based JSCC model is the Attention-based Deep JSCC (ADJSCC)\cite{yang_deep_2021}. The ADJSCC model introduces an attention mechanism to encoding channel SNR, which enables it to effectively operate with SNR levels during transmission. In order to demonstrate the effectiveness of our proposed SIJSCC model, we conducted experiments to compare it with the ADJSCC model. The training method and parameters of the ADJSCC model were kept consistent with the original paper.
To ensure a fair comparison with ADJSCC, we maintained a fixed model transmission rate of $Ratio=1/6$ and $Ratio=1/12$, and evaluate the performance on the Kodak24 test set. This test set comprises of 24 images, each with a size of $768\times512$.
\begin{figure}
    \centering
    \subfloat[]{\includegraphics[width=\linewidth]{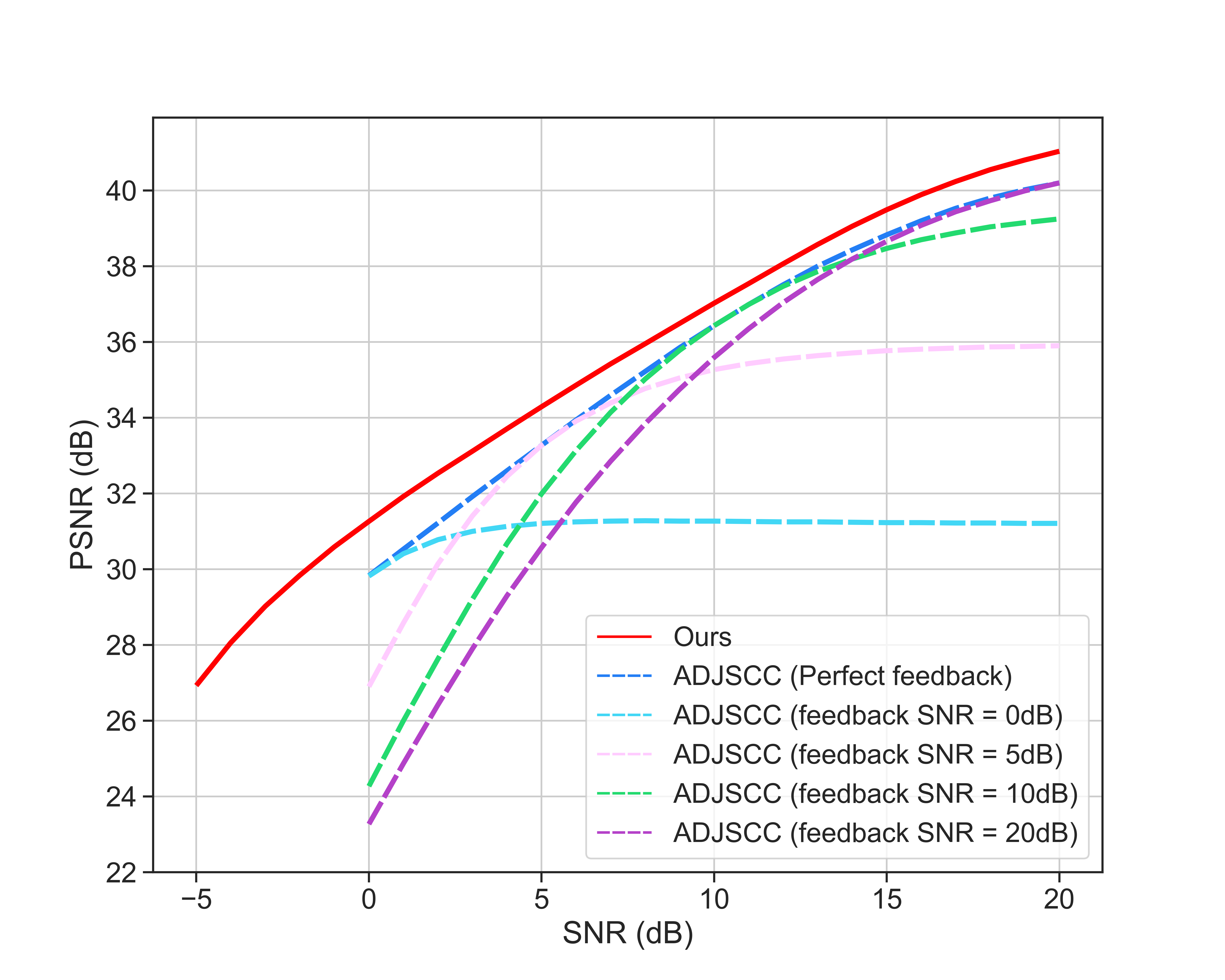}\label{fig:SIJSCC_vs_adjscc_1-6}}
    \hfil
    \subfloat[]{\includegraphics[width=\linewidth]{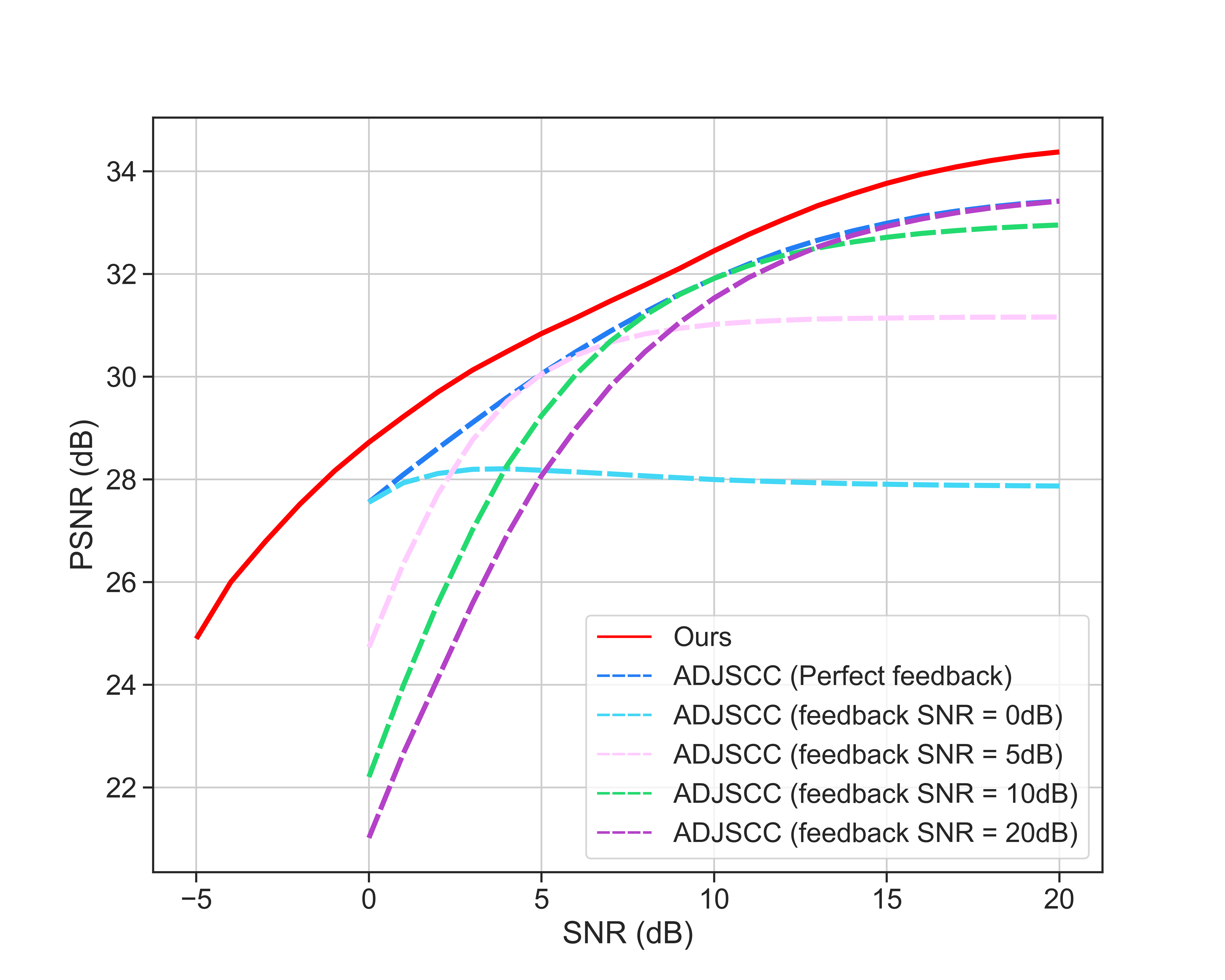}\label{fig:SIJSCC_vs_adjscc_1-12}}
    \caption{Performance comparison of SIJSCC and ADJSCC on the Kodak dataset test set. (a) ratio=1/6 and (b) ratio=1/12. The solid red curve represents the performance of our proposed method at multiple SNR, and the dashed line represents the performance of ADJSCC with different SNR feedback conditions.}
    \label{fig:SIJSCC_vs_adjscc}
\end{figure}

The final result is shown in Fig. \ref{fig:SIJSCC_vs_adjscc}, where the image reconstruction quality of both SIJSCC and ADJSCC models were presented across various SNR levels. Firstly, we only consider ADJSCC with perfect SNR feedback. Both SIJSCC and ADJSCC demonstrate the ability to adapt to changing SNR and exhibit smooth descent curves as the SNR decreases. However, the proposed SIJSCC method demonstrates a clear advantage over ADJSCC when the compression rate is set to $1/6$, as shown in Fig. \ref{fig:SIJSCC_vs_adjscc_1-6}, across all levels of SNR. This conclusion is not affected when the transmission rate is changed.
As shown in Figure \ref{fig:SIJSCC_vs_adjscc_1-12}, as the transmission rate is reduced from $1/6$ to $1/12$, SIJSCC continues to maintain a significant advantage over ADJSCC, with a lead of approximately 1dB PSNR across all SNR levels.

It is important to note that our current analysis is limited to ADJSCC with perfect SNR feedback. The dashed line in Figure 4 represents the reconstruction performance of ADJSCC in the presence of channel mismatch. As the mismatch error increases, the reconstruction performance of ADJSCC significantly decreases, particularly at low SNR level. This indicates that in practical applications, the performance difference between the two methods will be even more pronounced. In contrast, SIJSCC does not rely on channel information during the encoding and decoding process, and thus does not suffer from performance loss due to channel mismatch.

\begin{figure*}
    \centering
    \subfloat[]{\includegraphics[width=\linewidth]{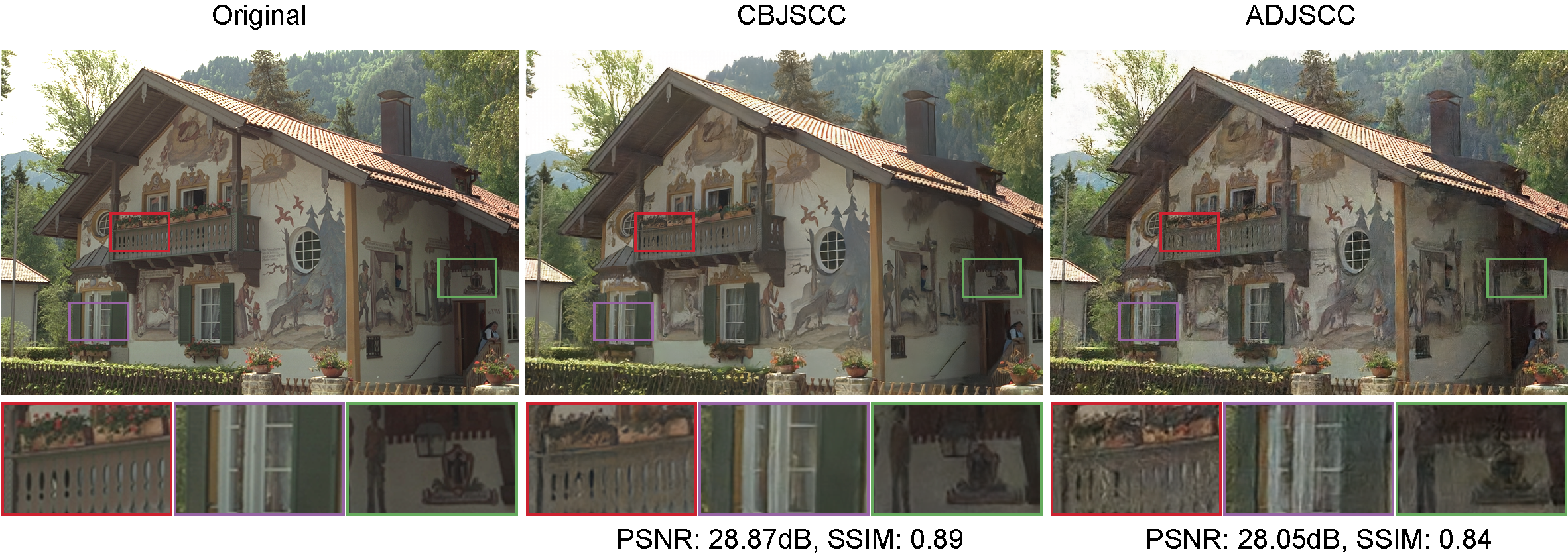}\label{fig:kodim24-1dB}}
    \hfil
    \subfloat[]{\includegraphics[width=\linewidth]{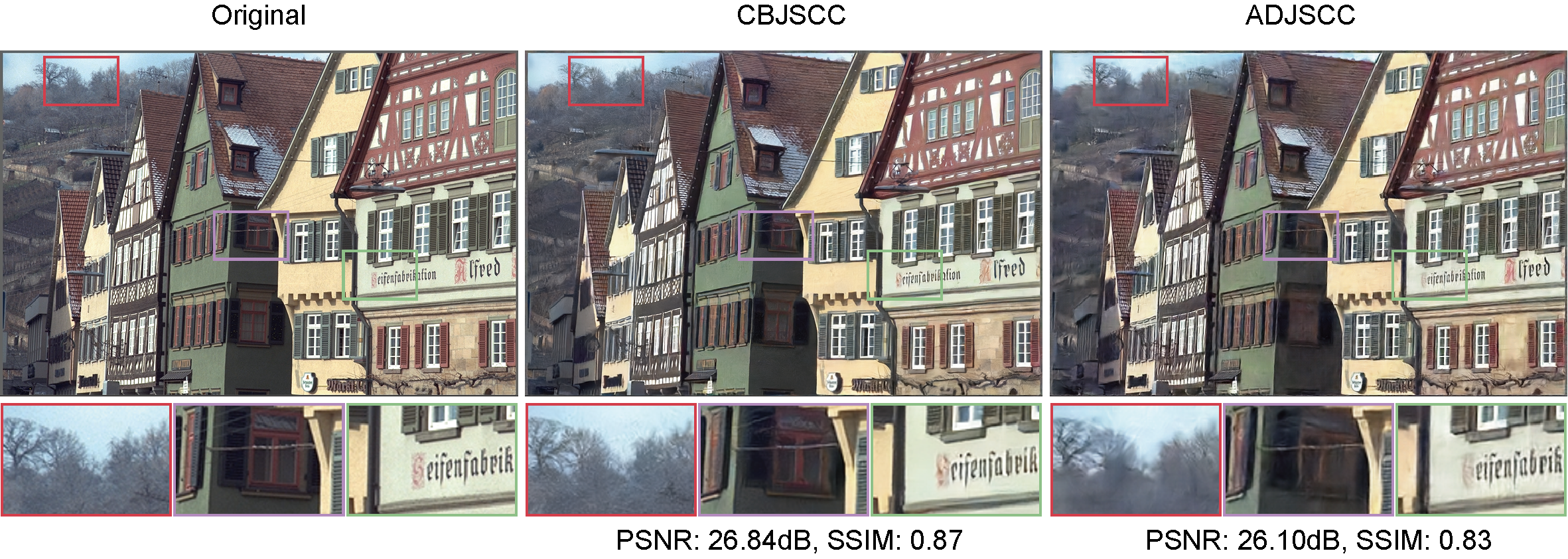}\label{fig:kodim08-1dB}}
    \caption{The performance of SIJSCC and ADJSCC in reconstructing images under an AWGN channel with an SNR of 1dB and different transmission rates from the Kodak dataset, using rates of (a) $1/6$ and (b) $1/12$, is shown in the images. Our proposed method's reconstructed images are shown in the middle, while those reconstructed using ADJSCC are shown on the right. The cropped image corresponding to the box is shown below the image.}
    \label{fig:kodim_visulize}
\end{figure*}

We have provided a visual comparison between SIJSCC and ADJSCC for images from the Kodak dataset in Figure \ref{fig:kodim_visulize}. Here, we only consider perfect SNR feedback for ADJSCC. SIJSCC and ADJSCC both exhibit excellent reconstruction ability overall, however, the performance of ADJSCC is still unsatisfactory at the edges of the image. By comparing the enlarged boxes in Fig. \ref{fig:kodim24-1dB}, we can observe that SIJSCC can more accurately restore the details and structures of the original image, while ADJSCC exhibits some blurring and distortion. In Fig. \ref{fig:kodim08-1dB}, some distant branches within the red box appear blurry and lose many details in the image reconstructed by ADJSCC. However, ADJSCC preserves the clear texture of these branches and produces a more realistic image. Based on this, it can be concluded that SIJSCC is superior to ADJSCC, particularly in high-frequency regions that exhibit rich variations.

\subsection{Impact of SNR information}
\begin{figure*}
    \centering
    \subfloat[]{\includegraphics[height=3in]{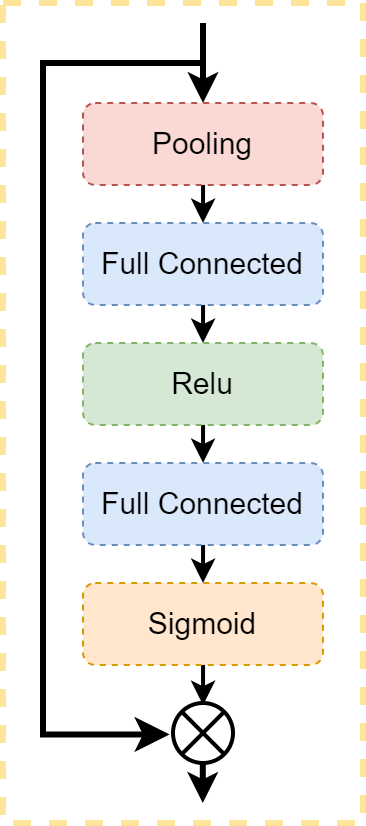}\label{fig:af_module}}
    \subfloat[]{\includegraphics[height=3in]{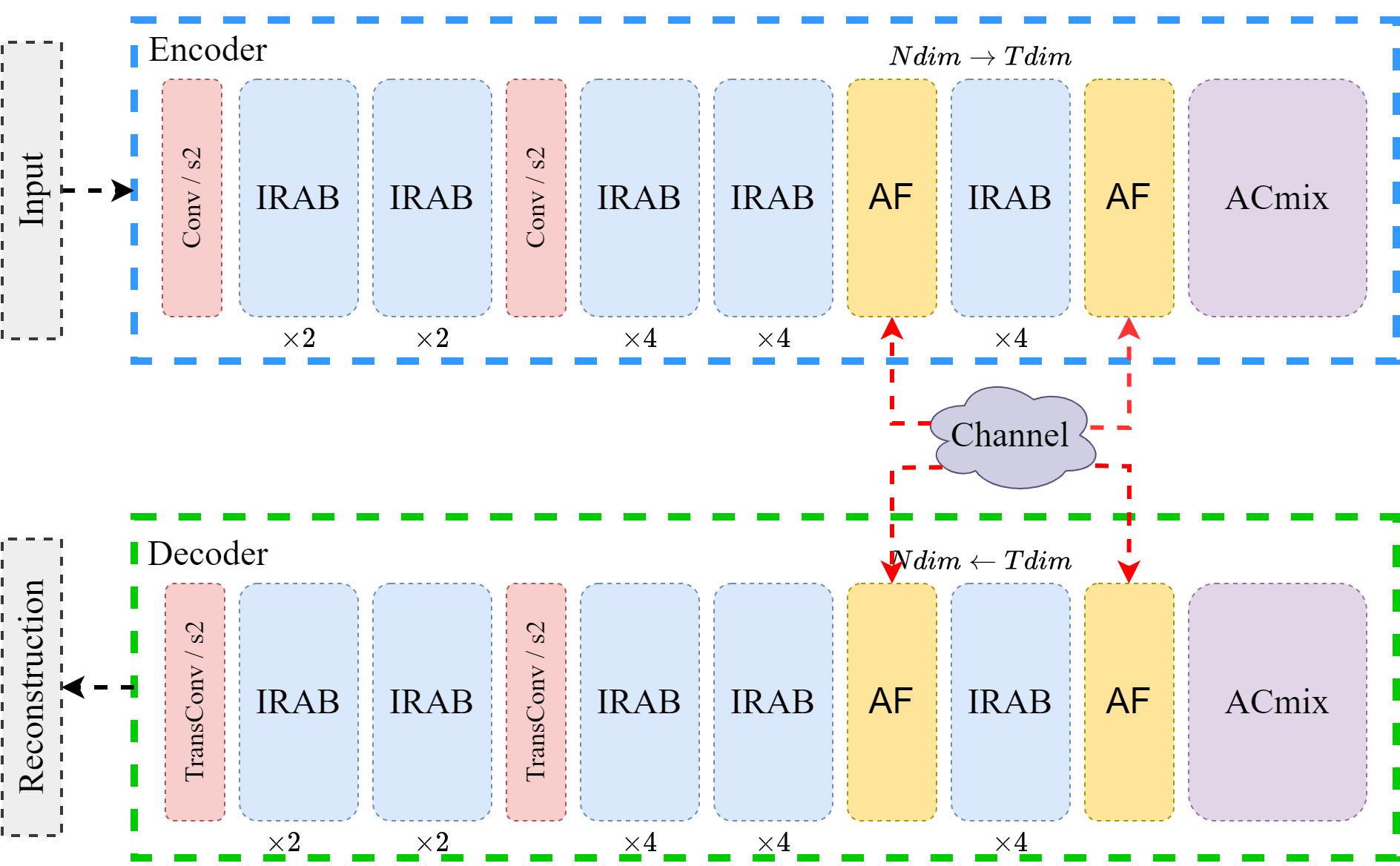}\label{fig:channel-prior-analysis}}
    \caption{(a) The architecture of the AF module in ADJSCC. (b) A modified SIJSCC with the AF module. The red dashed line represents the input of the SNR information. The yellow block represents the inserted AF module.}
\end{figure*}

Previous experiments have shown that SIJSCC outperforms ADJSCC without relying on SNR information, which implies that neutral networks may be able to adapt to the characteristics of the channel.
To further analyze the impact of channel feedback on the learning-based joint encoding and decoding model, we inserted the Attention Feature (AF) module used in ADJSCC into the encoder and decoder of SIJSCC, as shown in Fig. \ref{fig:af_module}.
The AF module can be seen as a type of channel attention mechanism that encodes channel SNR information into the network.
We have added two AF modules in both the encoder and decoder as in Fig. \ref{fig:channel-prior-analysis}, and we believe that this is sufficient to allow the network to adjust based on channel information.

We will now discuss three different conditions. The first condition assumes that both the encoder and decoder have access to the channel state information, which is similar to ADJSCC. The second scenario assumes that only the decoder has access to the channel state information and channel feedback is not available. The third scenario assumes that neither the encoder nor the decoder has access to the channel state information, which is proposed in this study.

\begin{figure}
    \centering
    \includegraphics[width=\linewidth]{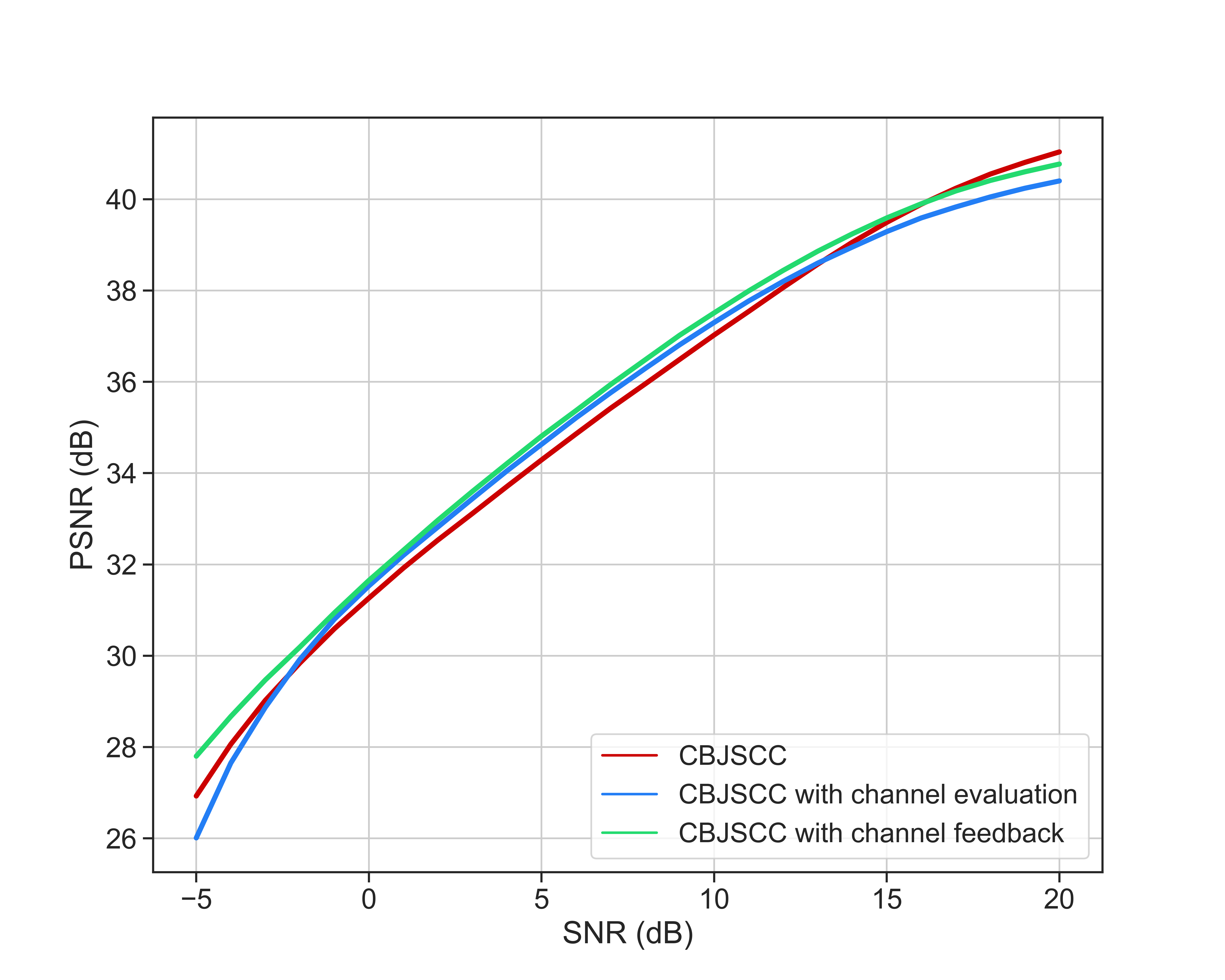}
    \caption{PSNR curves for reconstructed images under three different scenarios.}
    \label{fig:channel_prior_result}
\end{figure}

The result is shown in Fig. \ref{fig:channel_prior_result}, when the channel SNR is -5dB, SNR feedback performs better than others. However, this advantage is not significant, and as the channel SNR increases, the difference between the methods disappears. This could be due to the uncertainty in the deep learning model. For instance, when the SNR is equal to 20dB, the model without SNR information performs better.

Goblick et al.'s derivation \cite{goblick_theoretical_1965} shows that setting the source rate to the channel capacity can achieve minimum distortion for a given SNR ratio.  However, our experimental results showed that the proposed model did not significantly benefit from channel estimation and channel feedback.
A possible explanation is that the proposed method utilize the powerful fitting ability of neural networks, which can learn the characteristics of the channel and potential encoding and decoding methods from data, thereby achieving minimum distortion without estimating channel capacity. This means that deep learning models can automatically adjust the parameters of encoding and decoding by learning the statistical characteristics of the data, in order to minimize distortion during the transmission process. This has important implications for studying new joint encoding and decoding methods.
However, due to the fact that the deep learning model is a black box, and the internal mechanism cannot be directly observed, we maintain an open attitude towards this conclusion.

\subsection{Effective Model Complexity}

The previous results have demonstrated the powerful feature extraction and reconstruction capabilities of SIJSCC without relying on SNR information, which can be attributed to the carefully designed network structure of SIJSCC.
Here, we further analyze the relationship between expressive capacity and effective model complexity in SIJSCC. Expressive capacity represent the capacity of deep learning models in approximating complex problems, while effective model complexity measures the usable capacity of a trained model and reflects its complexity in practical JSCC tasks. The expressive capacity can be regarded as the upper bound of the amount of knowledge that a model architecture can hold\cite{hu_model_2021}. Exploring model complexity is crucial not only for comprehending the inner workings of a deep model but also for investigating the fundamental questions related to JSCC. The capacity of the model can be controlled by adjusting the value of $N$ in IRAB block as in Fig. \ref{fig:SIJSCC} ,while keeping the transmission dimension $T$ unchanged.
\begin{table*}
    \begin{center}
        \caption{EMC analysis across different model capacity}\label{tab:model_capacity}
        \begin{tabular}{|l|c|c|c|c|c|c|c|}
            \hline
            \multirow{2}{*}{\textbf{Model(N dim)}} &
            \multirow{2}{*}{\textbf{Params(M)}}    &
            \multirow{2}{*}{\textbf{MACs(G)}}      &
            \multicolumn{5}{c|}{\textbf{PSNR(dB)}}                                                                                                                                                              \\ \cline{4-8}
                                                   &       &        & \textbf{SNR1dB}           & \textbf{SNR4dB}           & \textbf{SNR7dB}           & \textbf{SNR13dB}          & \textbf{SNR19dB}
            \\ \hline
            SIJSCC(64)                             & 0.87  & 48.10  & 28.53                     & 29.36                     & 29.90                     & 30.04                     & 30.54                     \\
            SIJSCC(96)                             & 1.85  & 104.26 & 30.55                     & 32.39                     & 34.06                     & 36.55                     & 37.72                     \\
            SIJSCC(128)                            & 3.21  & 182.05 & 30.95                     & 32.88                     & 34.71                     & 37.85                     & 39.72                     \\
            SIJSCC(160)                            & 4.96  & 281.49 & 31.30                     & 33.14                     & 34.90                     & 37.96                     & 39.76                     \\
            SIJSCC(192)                            & 7.09  & 402.57 & \color{red}\textbf{31.90} & \color{red}\textbf{33.71} & \color{blue}35.41         & \color{blue}38.58         & \color{blue}40.81         \\
            SIJSCC(256)                            & 12.49 & 709.66 & \color{blue}31.81         & \color{blue}33.66         & \color{red}\textbf{35.45} & \color{red}\textbf{38.70} & \color{red}\textbf{40.97} \\
            ADJSCC                                 & 10.76 & 398.18 & 30.53                     & 32.60                     & 34.59                     & 38.00                     & 40.02                     \\
            \hline
        \end{tabular}
    \end{center}
\end{table*}
Here, we evaluate the models using the Kodak dataset, and the transmission rate is fixed at $1/6$.
As presented in the Tab. \ref{tab:model_capacity}, we evaluated the performance of SIJSCC at 5 different values of $N$. We also compared the performance with that of ADJSCC. Multiply-Accumulate Operations (MACs) and Parameters are widely used indicators for evaluating model complexity. The MACs are calculated based on the image size of $768\times 512$ in the Kodak dataset. The red values represent the top-1 PSNR result under the current channel, while the blue values represent the top-2 result.

It is evident the performance of SIJSCC gradually improves with increasing $N$. However, when $N$ is set to 192, the MACs of SIJSCC are similar to those of ADJSCC, but the performance of SIJSCC is significantly better. This indicates that the performance improvement of SIJSCC compared to ADJSCC is mainly due to the improvement of the network structure, rather than an increase in expressive capacity.
On the other hand, we can observe that when $N=128$, the performance of SIJSCC does not show a significant decrease compared when $N$ is 192. In fact, it is comparable to the performance of ADJSCC with perfect feedback. However, the number of parameters is only one-third of that of ADJSCC, and the MACs are only 182.05G, which is about half of the default SIJSCC. This indicates that the network structure of SIJSCC is designed to be very compact, which can reduce the number of parameters while ensuring performance. This is very meaningful for practical deployment.

The top-1 and top-2 PSNR results are distributed similarly between $N=128$ and $N=256$, with values that are very close. However, the MACs and parameters of the model at $N=256$ are nearly twice that of $N=192$. Additionally, SIJSCC achieves higher PSNR at $N=192$ in low SNR conditions, indicating that the performance of SIJSCC when $N=192$ is close to the upper bound of its expressive capacity.
We also observed that SIJSCC at $N=160$ only maintains a slight advantage over $N=128$ when the SNR is less than 5dB. Therefore, to achieve a balance between performance and resource consumption, we can choose a suitable model based on the communication environment. Additionally, we can further explore the design of a more lightweight SIJSCC network.

\subsection{Domain Adaptation Ability}

In the field of deep learning, the process of training a model on a dataset from one domain and testing it on a dataset from a different domain is referred to as "domain adaptation" or "domain transfer". This section aims to evaluate the robustness and adaptability of SIJSCC to different data domains This process is challenging as the training dataset may not encompass all the real-world scenarios, leading to a significant decrease in performance in practical applications. Our objective is to develop a model that can generalize well across different domains and prevent over fitting to the dataset, which can result in poor performance on unseen data.

\begin{figure}
    \centering
    \includegraphics[width=\linewidth]{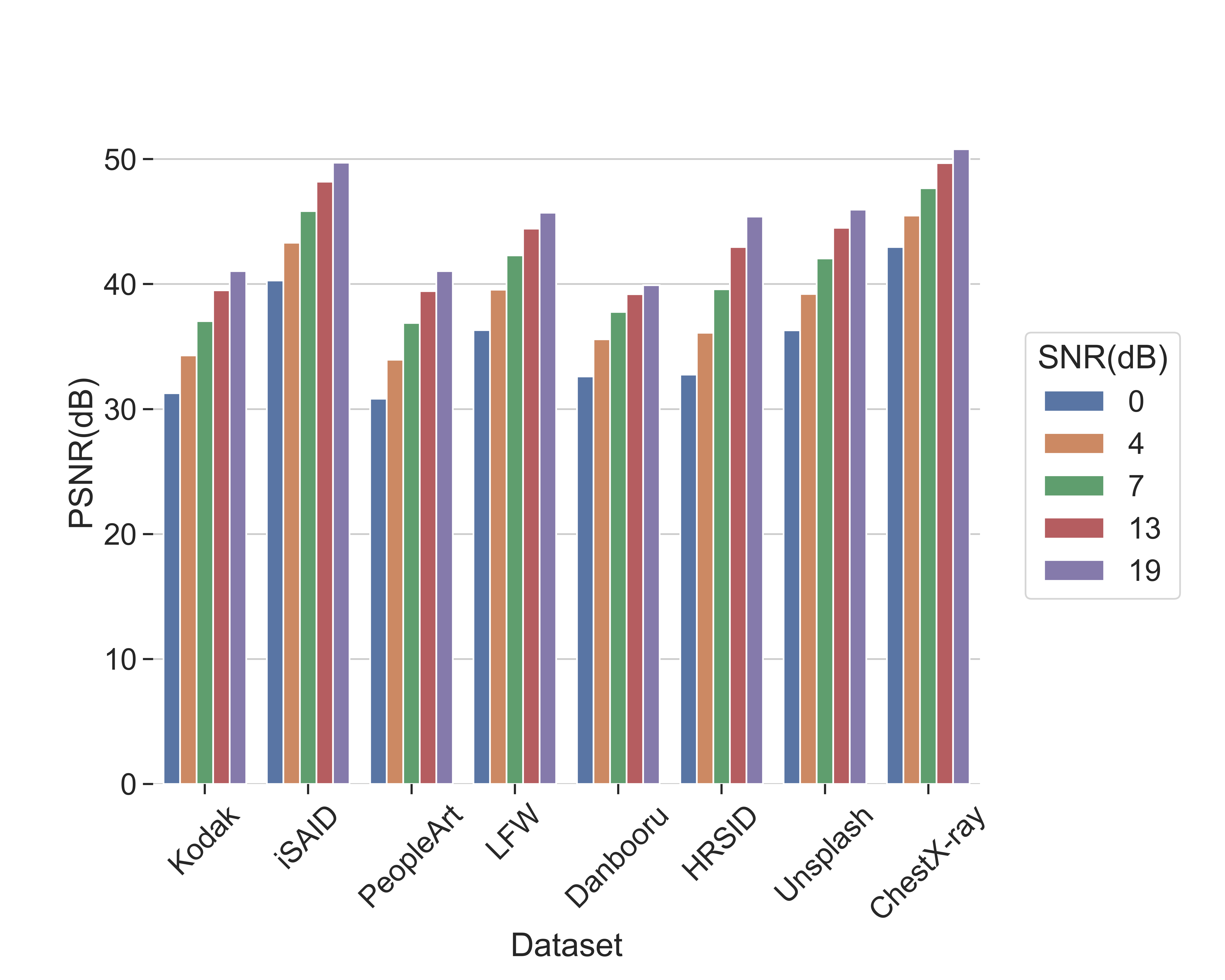}
    \caption{Clustered Column Chart of SIJSCC under multiple datasets domain, the color of the line represents channel SNR.}
    \label{fig:domain_adaptation}
\end{figure}

As mentioned previously, SIJSCC was trained on the ImageNet dataset. Here we validated the trained model using selected test datasets that can represent a variety of scenarios, as presented in Fig. \ref{fig:domain_adaptation}. iSAID\cite{lai_deep_2017} is a large-scale dataset for instance segmentation of aerial images with high spatial resolution. People-Art\cite{charbonnier_two_1994} is an object detection dataset composed of 43 different styles of people. The characters depicted in this dataset are quite different from those in ordinary photographs. Labeled Faces in the Wild (LFW)\cite{russakovsky_imagenet_2015} contains over 13,000 internet-downloaded face images with considerable variations in pose, illumination, and facial expression, making it one of the most well-known face recognition datasets in the field of computer vision. Danbooru2021\cite{lim_enhanced_2017} is a large anime-style image database with 4.9 million illustrations or anime-style images. The High-Resolution SAR Images Dataset (HRSID)\cite{weiHRSIDHighResolutionSAR2020} is a dataset designed for ship detection, semantic segmentation, and instance segmentation tasks in high-resolution Synthetic Aperture Radar (SAR) images, including SAR images with different resolutions, polarizations, sea states, coastal zones, and ports. The Unsplash Dataset\cite{UnsplashDatasetWorld2023} includes a vast collection of high-quality images contributed by professional photographers from all around the world, covering a wide range of camera brands, models, lenses, and focal lengths. ChestX-ray8 \cite{wang_chestx-ray8_nodate} is a dataset from the medical field that contains a large number of X-ray images.

The final results indicate that SIJSCC performs consistently well on benchmark datasets across various fields. Across all datasets, the PSNR of reconstructed image exceeded 30dB with an SNR level of 0dB. With the exception of the Danbooru2021 dataset, nearly all datasets achieved a PSNR exceeding 40dB when SNR equal 19dB. It is noteworthy that the model achieved a reconstruction performance of PSNR 50dB on the CT Medical dataset at a 19dB SNR level, even though the training dataset did not include medical CT image data. This could be attributed to the fact that CT images have smoother color and texture variations and contain fewer high-frequency details. Similarly, SIJSCC also demonstrated excellent performance on the iSAID dataset, which contains a large number of water areas with small frequency changes. Danbooru2021 and PeopleArt belong to the domains of art and cartoon images, which typically exhibit strong artistic styles and visual effects, such as bright colors, unique compositions, and smooth lines. While the proposed SIJSCC model demonstrated slightly lower reconstruction performance on these two datasets, it still achieved a reconstruction performance of over 30dB PSNR at an SNR level of 0dB.

In summary, the experimental results demonstrate that the proposed SIJSCC model achieves high reconstruction performance on benchmark datasets across various domains, including some unseen domain. These results suggest that the SIJSCC model is a promising approach for image JSCC task.

\section{Conclusion}\label{sec:conclusion}
This article proposes a DL-based joint source-channel coding method, SIJSCC, which can adapt channel automatically without relying on SNR information.
Compared to existing DL-based JSCC methods based on channel feedback, the proposed method achieves better performance.
This improvement is attributed to the carefully designed encoder-decoder and the integration of attention mechanisms. We conducted experiments to investigate the effect of SNR information on DL-based JSCC tasks and found no significant improvement.
This seems to suggest that encoder and decoder based on neural networks differ from traditional methods in that they can autonomously learn the potential patterns of the channel, thereby reducing the impact of channel noise and other channel errors.
However, we need to exercise caution in drawing such a conclusion, given the complexity of deep learning.
This idea can also be further extended to other time-varying channels, further research is needed to gain a better understanding of the role of channel information in DL-based JSCC.
The experimental results indicate that our method exhibits good scalability and robustness on data from multiple domains. It could be a promising approach in situations where channel models are unknown or difficult to model analytically, such as in high mobility vehicular communications, unmanned Aerial Vehicle communication, deep space communication and underwater communications, among other scenarios.

\bibliographystyle{IEEEtran}
\bibliography{references}

\begin{thebibliography}{10}
\providecommand{\url}[1]{#1}
\csname url@samestyle\endcsname
\providecommand{\newblock}{\relax}
\providecommand{\bibinfo}[2]{#2}
\providecommand{\BIBentrySTDinterwordspacing}{\spaceskip=0pt\relax}
\providecommand{\BIBentryALTinterwordstretchfactor}{4}
\providecommand{\BIBentryALTinterwordspacing}{\spaceskip=\fontdimen2\font plus
\BIBentryALTinterwordstretchfactor\fontdimen3\font minus
  \fontdimen4\font\relax}
\providecommand{\BIBforeignlanguage}[2]{{%
\expandafter\ifx\csname l@#1\endcsname\relax
\typeout{** WARNING: IEEEtran.bst: No hyphenation pattern has been}%
\typeout{** loaded for the language `#1'. Using the pattern for}%
\typeout{** the default language instead.}%
\else
\language=\csname l@#1\endcsname
\fi
#2}}
\providecommand{\BIBdecl}{\relax}
\BIBdecl

\bibitem{gastpar_code_2003}
M.~Gastpar, B.~Rimoldi, and M.~Vetterli, ``To code, or not to code: lossy
  source-channel communication revisited,'' \emph{IEEE Transactions on
  Information Theory}, vol.~49, no.~5, pp. 1147--1158, May 2003, conference
  Name: IEEE Transactions on Information Theory.

\bibitem{kostina_joint_2017}
V.~Kostina, Y.~Polyanskiy, and S.~Verd, ``Joint {Source}-{Channel} {Coding}
  {With} {Feedback},'' \emph{IEEE Transactions on Information Theory}, vol.~63,
  no.~6, pp. 3502--3515, June 2017, conference Name: IEEE Transactions on
  Information Theory.

\bibitem{skoglund_design_2002}
M.~Skoglund, N.~Phamdo, and F.~Alajaji, ``Design and performance of {VQ}-based
  hybrid digital-analog joint source-channel codes,'' \emph{IEEE Transactions
  on Information Theory}, vol.~48, no.~3, pp. 708--720, March 2002, conference
  Name: IEEE Transactions on Information Theory.

\bibitem{mittal_hybrid_2002}
U.~Mittal and N.~Phamdo, ``Hybrid digital-analog ({HDA}) joint source-channel
  codes for broadcasting and robust communications,'' \emph{IEEE Transactions
  on Information Theory}, vol.~48, no.~5, pp. 1082--1102, May 2002, conference
  Name: IEEE Transactions on Information Theory.

\bibitem{chatellier_robust_2007}
\BIBentryALTinterwordspacing
C.~Chatellier, H.~Boeglen, C.~Perrine, C.~Olivier, and O.~Haeberlé, ``A robust
  joint source channel coding scheme for image transmission over the
  ionospheric channel,'' \emph{Signal Processing: Image Communication},
  vol.~22, no. 2007, pp. 543--556, May 2007, publisher: Elsevier. [Online].
  Available: \url{https://hal.archives-ouvertes.fr/hal-00431283}
\BIBentrySTDinterwordspacing

\bibitem{kozintsev_robust_1998}
I.~Kozintsev and K.~Ramchandran, ``Robust image transmission over
  energy-constrained time-varying channels using multiresolution joint
  source-channel coding,'' \emph{IEEE Transactions on Signal Processing},
  vol.~46, no.~4, pp. 1012--1026, April 1998, conference Name: IEEE
  Transactions on Signal Processing.

\bibitem{yang_introduction_2022}
\BIBentryALTinterwordspacing
Y.~Yang, S.~Mandt, and L.~Theis, ``An {Introduction} to {Neural} {Data}
  {Compression},'' November 2022, arXiv:2202.06533 [cs, eess, math]. [Online].
  Available: \url{http://arxiv.org/abs/2202.06533}
\BIBentrySTDinterwordspacing

\bibitem{xu_wireless_2022}
J.~Xu, B.~Ai, W.~Chen, A.~Yang, P.~Sun, and M.~Rodrigues, ``Wireless {Image}
  {Transmission} {Using} {Deep} {Source} {Channel} {Coding} {With} {Attention}
  {Modules},'' \emph{IEEE Transactions on Circuits and Systems for Video
  Technology}, vol.~32, no.~4, pp. 2315--2328, April 2022, conference Name:
  IEEE Transactions on Circuits and Systems for Video Technology.

\bibitem{yang_deep_2021}
M.~Yang, C.~Bian, and H.-S. Kim, ``Deep {Joint} {Source} {Channel} {Coding} for
  {Wireless} {Image} {Transmission} with {OFDM},'' in \emph{{ICC} 2021 - {IEEE}
  {International} {Conference} on {Communications}}, June 2021, pp. 1--6, iSSN:
  1938-1883.

\bibitem{wang_novel_2021}
S.~Wang, J.~Dai, S.~Yao, K.~Niu, and P.~Zhang, ``A {Novel} {Deep} {Learning}
  {Architecture} for {Wireless} {Image} {Transmission},'' in \emph{2021 {IEEE}
  {Global} {Communications} {Conference} ({GLOBECOM})}, December 2021, pp.
  1--6.

\bibitem{burth_kurka_joint_2020}
D.~Burth~Kurka and D.~G{\"u}nd{\"u}z, ``Joint source-channel coding of images
  with (not very) deep learning,'' in \emph{International Zurich Seminar on
  Information and Communication (IZS 2020). Proceedings}.\hskip 1em plus 0.5em
  minus 0.4em\relax ETH Zurich, 2020, pp. 90--94.

\bibitem{bourtsoulatze_deep_2019}
E.~Bourtsoulatze, D.~Burth~Kurka, and D.~Gündüz, ``Deep {Joint}
  {Source}-{Channel} {Coding} for {Wireless} {Image} {Transmission},''
  \emph{IEEE Transactions on Cognitive Communications and Networking}, vol.~5,
  no.~3, pp. 567--579, September 2019, 85 citations (Crossref) [2022-10-04]
  Conference Name: IEEE Transactions on Cognitive Communications and
  Networking.

\bibitem{kurka_deepjscc-f_2020}
D.~B. Kurka and D.~Gündüz, ``{DeepJSCC}-f: {Deep} {Joint} {Source}-{Channel}
  {Coding} of {Images} {With} {Feedback},'' \emph{IEEE Journal on Selected
  Areas in Information Theory}, vol.~1, no.~1, pp. 178--193, May 2020,
  conference Name: IEEE Journal on Selected Areas in Information Theory.

\bibitem{yang_deep_2022}
M.~Yang and H.-S. Kim, ``Deep {Joint} {Source}-{Channel} {Coding} for
  {Wireless} {Image} {Transmission} with {Adaptive} {Rate} {Control},'' in
  \emph{{ICASSP} 2022 - 2022 {IEEE} {International} {Conference} on
  {Acoustics}, {Speech} and {Signal} {Processing} ({ICASSP})}, May 2022, pp.
  5193--5197, iSSN: 2379-190X.

\bibitem{sun_joint_2022}
L.~Sun, C.~Guo, and Y.~Yang, ``Joint {Source}-{Channel} {Coding} for
  {Efficient} {Image} {Transmission}: {An} {Information} {Bottleneck} {Based}
  {Scheme},'' in \emph{2022 {IEEE} {Globecom} {Workshops} ({GC} {Wkshps})},
  December 2022, pp. 1852--1857.

\bibitem{ding_snr-adaptive_2021}
M.~Ding, J.~Li, M.~Ma, and X.~Fan, ``{SNR}-{Adaptive} {Deep} {Joint}
  {Source}-{Channel} {Coding} for {Wireless} {Image} {Transmission},'' in
  \emph{{ICASSP} 2021 - 2021 {IEEE} {International} {Conference} on
  {Acoustics}, {Speech} and {Signal} {Processing} ({ICASSP})}, June 2021, pp.
  1555--1559, iSSN: 2379-190X.

\bibitem{simonyan_very_2015}
\BIBentryALTinterwordspacing
K.~Simonyan and A.~Zisserman, ``\BIBforeignlanguage{en}{Very {Deep}
  {Convolutional} {Networks} for {Large}-{Scale} {Image} {Recognition}},''
  \emph{\BIBforeignlanguage{en}{arXiv:1409.1556 [cs]}}, April 2015, arXiv:
  1409.1556. [Online]. Available: \url{http://arxiv.org/abs/1409.1556}
\BIBentrySTDinterwordspacing

\bibitem{yang2022resolution}
K.~Yang, S.~Wang, K.~Tan, J.~Dai, D.~Zhou, and K.~Niu, ``Resolution-adaptive
  source-channel coding for end-to-end wireless image transmission,'' in
  \emph{GLOBECOM 2022-2022 IEEE Global Communications Conference}.\hskip 1em
  plus 0.5em minus 0.4em\relax IEEE, 2022, pp. 1460--1465.

\bibitem{liu_learned_2021}
\BIBentryALTinterwordspacing
C.~Liu, H.~Sun, J.~Katto, X.~Zeng, and Y.~Fan, ``Learned {Video} {Compression}
  with {Residual} {Prediction} and {Loop} {Filter},'' \emph{arXiv:2108.08551
  [cs, eess]}, August 2021, arXiv: 2108.08551 version: 1. [Online]. Available:
  \url{http://arxiv.org/abs/2108.08551}
\BIBentrySTDinterwordspacing

\bibitem{bao_adjscc-l_2021}
X.~Bao, M.~Jiang, and H.~Zhang, ``{ADJSCC}-l: {SNR}-{Adaptive} {JSCC}
  {Networks} for {Multi}-{Layer} {Wireless} {Image} {Transmission},'' in
  \emph{2021 7th {International} {Conference} on {Computer} and
  {Communications} ({ICCC})}, December 2021, pp. 1812--1816.

\bibitem{wu_channel-adaptive_2022}
H.~Wu, Y.~Shao, K.~Mikolajczyk, and D.~Gündüz, ``Channel-{Adaptive}
  {Wireless} {Image} {Transmission} {With} {OFDM},'' \emph{IEEE Wireless
  Communications Letters}, vol.~11, no.~11, pp. 2400--2404, November 2022,
  conference Name: IEEE Wireless Communications Letters.

\bibitem{kim_deepcode_2020}
H.~Kim, Y.~Jiang, S.~Kannan, S.~Oh, and P.~Viswanath, ``Deepcode: {Feedback}
  {Codes} via {Deep} {Learning},'' \emph{IEEE Journal on Selected Areas in
  Information Theory}, vol.~1, no.~1, pp. 194--206, May 2020, conference Name:
  IEEE Journal on Selected Areas in Information Theory.

\bibitem{shannon_zero_1956}
C.~Shannon, ``The zero error capacity of a noisy channel,'' \emph{IRE
  Transactions on Information Theory}, vol.~2, no.~3, pp. 8--19, September
  1956, conference Name: IRE Transactions on Information Theory.

\bibitem{woo_convnext_2023}
\BIBentryALTinterwordspacing
S.~Woo, S.~Debnath, R.~Hu, X.~Chen, Z.~Liu, I.~S. Kweon, and S.~Xie,
  ``{ConvNeXt} {V2}: {Co}-designing and {Scaling} {ConvNets} with {Masked}
  {Autoencoders},'' January 2023, arXiv:2301.00808 [cs]. [Online]. Available:
  \url{http://arxiv.org/abs/2301.00808}
\BIBentrySTDinterwordspacing

\bibitem{liu_residual_2020}
\BIBentryALTinterwordspacing
J.~Liu, W.~Zhang, Y.~Tang, J.~Tang, and G.~Wu,
  ``\BIBforeignlanguage{en}{Residual {Feature} {Aggregation} {Network} for
  {Image} {Super}-{Resolution}},'' in \emph{\BIBforeignlanguage{en}{2020
  {IEEE}/{CVF} {Conference} on {Computer} {Vision} and {Pattern} {Recognition}
  ({CVPR})}}.\hskip 1em plus 0.5em minus 0.4em\relax Seattle, WA, USA: IEEE,
  June 2020, pp. 2356--2365. [Online]. Available:
  \url{https://ieeexplore.ieee.org/document/9156371/}
\BIBentrySTDinterwordspacing

\bibitem{kong_residual_2022}
\BIBentryALTinterwordspacing
F.~Kong, M.~Li, S.~Liu, D.~Liu, J.~He, Y.~Bai, F.~Chen, and L.~Fu, ``Residual
  {Local} {Feature} {Network} for {Efficient} {Super}-{Resolution},'' May 2022,
  arXiv:2205.07514 [cs]. [Online]. Available:
  \url{http://arxiv.org/abs/2205.07514}
\BIBentrySTDinterwordspacing

\bibitem{pan_integration_2022}
\BIBentryALTinterwordspacing
X.~Pan, C.~Ge, R.~Lu, S.~Song, G.~Chen, Z.~Huang, and G.~Huang, ``On the
  {Integration} of {Self}-{Attention} and {Convolution},'' March 2022,
  arXiv:2111.14556 [cs]. [Online]. Available:
  \url{http://arxiv.org/abs/2111.14556}
\BIBentrySTDinterwordspacing

\bibitem{rumelhart_learning_1988}
\BIBentryALTinterwordspacing
D.~Rumelhart, G.~Hinton, and R.~Williams, ``\BIBforeignlanguage{en}{Learning
  {Internal} {Representations} by {Error} {Propagation}},'' in
  \emph{\BIBforeignlanguage{en}{Readings in {Cognitive} {Science}}}.\hskip 1em
  plus 0.5em minus 0.4em\relax Elsevier, 1988, pp. 399--421. [Online].
  Available:
  \url{https://linkinghub.elsevier.com/retrieve/pii/B9781483214467500352}
\BIBentrySTDinterwordspacing

\bibitem{hinton_reducing_2006}
\BIBentryALTinterwordspacing
G.~E. Hinton and R.~R. Salakhutdinov, ``Reducing the {Dimensionality} of {Data}
  with {Neural} {Networks},'' \emph{Science}, vol. 313, no. 5786, pp. 504--507,
  July 2006, publisher: American Association for the Advancement of Science.
  [Online]. Available:
  \url{https://www.science.org/doi/abs/10.1126/science.1127647}
\BIBentrySTDinterwordspacing

\bibitem{baldi_autoencoders_2012}
\BIBentryALTinterwordspacing
P.~Baldi, ``\BIBforeignlanguage{en}{Autoencoders, {Unsupervised} {Learning},
  and {Deep} {Architectures}},'' in \emph{\BIBforeignlanguage{en}{Proceedings
  of {ICML} {Workshop} on {Unsupervised} and {Transfer} {Learning}}}.\hskip 1em
  plus 0.5em minus 0.4em\relax JMLR Workshop and Conference Proceedings, June
  2012, pp. 37--49, iSSN: 1938-7228. [Online]. Available:
  \url{https://proceedings.mlr.press/v27/baldi12a.html}
\BIBentrySTDinterwordspacing

\bibitem{balle_end--end_2017}
\BIBentryALTinterwordspacing
J.~Ballé, V.~Laparra, and E.~P. Simoncelli, ``End-to-end {Optimized} {Image}
  {Compression},'' March 2017, arXiv:1611.01704 [cs, math]. [Online].
  Available: \url{http://arxiv.org/abs/1611.01704}
\BIBentrySTDinterwordspacing

\bibitem{vincent_extracting_2008}
\BIBentryALTinterwordspacing
P.~Vincent, H.~Larochelle, Y.~Bengio, and P.-A. Manzagol, ``Extracting and
  composing robust features with denoising autoencoders,'' in \emph{Proceedings
  of the 25th international conference on {Machine} learning}, ser. {ICML}
  '08.\hskip 1em plus 0.5em minus 0.4em\relax New York, NY, USA: Association
  for Computing Machinery, July 2008, pp. 1096--1103. [Online]. Available:
  \url{https://doi.org/10.1145/1390156.1390294}
\BIBentrySTDinterwordspacing

\bibitem{hu_squeeze-and-excitation_nodate}
J.~Hu, L.~Shen, and G.~Sun, ``Squeeze-and-excitation networks,'' in
  \emph{Proceedings of the IEEE conference on computer vision and pattern
  recognition}, 2018, pp. 7132--7141.

\bibitem{park_bam_2018}
\BIBentryALTinterwordspacing
J.~Park, S.~Woo, J.-Y. Lee, and I.~S. Kweon, ``{BAM}: {Bottleneck} {Attention}
  {Module},'' July 2018, arXiv:1807.06514 [cs]. [Online]. Available:
  \url{http://arxiv.org/abs/1807.06514}
\BIBentrySTDinterwordspacing

\bibitem{hu_gather-excite_2019}
\BIBentryALTinterwordspacing
J.~Hu, L.~Shen, S.~Albanie, G.~Sun, and A.~Vedaldi, ``Gather-{Excite}:
  {Exploiting} {Feature} {Context} in {Convolutional} {Neural} {Networks},''
  January 2019, arXiv:1810.12348 [cs]. [Online]. Available:
  \url{http://arxiv.org/abs/1810.12348}
\BIBentrySTDinterwordspacing

\bibitem{woo_cbam_2018}
\BIBentryALTinterwordspacing
S.~Woo, J.~Park, J.-Y. Lee, and I.~S. Kweon, ``\BIBforeignlanguage{en}{{CBAM}:
  {Convolutional} {Block} {Attention} {Module}},''
  \emph{\BIBforeignlanguage{en}{arXiv:1807.06521 [cs]}}, July 2018, arXiv:
  1807.06521. [Online]. Available: \url{http://arxiv.org/abs/1807.06521}
\BIBentrySTDinterwordspacing

\bibitem{kim_mamnet_2018}
\BIBentryALTinterwordspacing
J.-H. Kim, J.-H. Choi, M.~Cheon, and J.-S. Lee, ``{MAMNet}: {Multi}-path
  {Adaptive} {Modulation} {Network} for {Image} {Super}-{Resolution},''
  November 2018, arXiv:1811.12043 [cs] version: 1. [Online]. Available:
  \url{http://arxiv.org/abs/1811.12043}
\BIBentrySTDinterwordspacing

\bibitem{vaswani_attention_2017}
\BIBentryALTinterwordspacing
A.~Vaswani, N.~Shazeer, N.~Parmar, J.~Uszkoreit, L.~Jones, A.~N. Gomez,
  L.~Kaiser, and I.~Polosukhin, ``\BIBforeignlanguage{en}{Attention {Is} {All}
  {You} {Need}},'' \emph{\BIBforeignlanguage{en}{arXiv:1706.03762 [cs]}},
  December 2017, arXiv: 1706.03762. [Online]. Available:
  \url{http://arxiv.org/abs/1706.03762}
\BIBentrySTDinterwordspacing

\bibitem{dosovitskiy_image_2021}
\BIBentryALTinterwordspacing
A.~Dosovitskiy, L.~Beyer, A.~Kolesnikov, D.~Weissenborn, X.~Zhai,
  T.~Unterthiner, M.~Dehghani, M.~Minderer, G.~Heigold, S.~Gelly, J.~Uszkoreit,
  and N.~Houlsby, ``An {Image} is {Worth} 16x16 {Words}: {Transformers} for
  {Image} {Recognition} at {Scale},'' \emph{arXiv:2010.11929 [cs]}, June 2021,
  arXiv: 2010.11929. [Online]. Available: \url{http://arxiv.org/abs/2010.11929}
\BIBentrySTDinterwordspacing

\bibitem{wang_non-local_2018}
\BIBentryALTinterwordspacing
X.~Wang, R.~Girshick, A.~Gupta, and K.~He, ``Non-local {Neural} {Networks},''
  April 2018, arXiv:1711.07971 [cs]. [Online]. Available:
  \url{http://arxiv.org/abs/1711.07971}
\BIBentrySTDinterwordspacing

\bibitem{liu_swin_2021}
Z.~Liu, Y.~Lin, Y.~Cao, H.~Hu, Y.~Wei, Z.~Zhang, S.~Lin, and B.~Guo, ``Swin
  transformer: Hierarchical vision transformer using shifted windows,'' in
  \emph{Proceedings of the IEEE/CVF international conference on computer
  vision}, 2021, pp. 10\,012--10\,022.

\bibitem{bello_attention_2019}
\BIBentryALTinterwordspacing
I.~Bello, B.~Zoph, Q.~Le, A.~Vaswani, and J.~Shlens,
  ``\BIBforeignlanguage{en}{Attention {Augmented} {Convolutional}
  {Networks}},'' in \emph{\BIBforeignlanguage{en}{2019 {IEEE}/{CVF}
  {International} {Conference} on {Computer} {Vision} ({ICCV})}}.\hskip 1em
  plus 0.5em minus 0.4em\relax Seoul, Korea (South): IEEE, October 2019, pp.
  3285--3294. [Online]. Available:
  \url{https://ieeexplore.ieee.org/document/9010285/}
\BIBentrySTDinterwordspacing

\bibitem{srinivas_bottleneck_2021}
A.~Srinivas, T.-Y. Lin, N.~Parmar, J.~Shlens, P.~Abbeel, and A.~Vaswani,
  ``Bottleneck transformers for visual recognition,'' in \emph{Proceedings of
  the IEEE/CVF conference on computer vision and pattern recognition}, 2021,
  pp. 16\,519--16\,529.

\bibitem{liu_convnet_2022}
\BIBentryALTinterwordspacing
Z.~Liu, H.~Mao, C.-Y. Wu, C.~Feichtenhofer, T.~Darrell, and S.~Xie, ``A
  {ConvNet} for the 2020s,'' March 2022, arXiv:2201.03545 [cs]. [Online].
  Available: \url{http://arxiv.org/abs/2201.03545}
\BIBentrySTDinterwordspacing

\bibitem{wang_residual_2017}
F.~Wang, M.~Jiang, C.~Qian, S.~Yang, C.~Li, H.~Zhang, X.~Wang, and X.~Tang,
  ``Residual attention network for image classification,'' in \emph{Proceedings
  of the IEEE conference on computer vision and pattern recognition}, 2017, pp.
  3156--3164.

\bibitem{balle_density_2016}
\BIBentryALTinterwordspacing
J.~Ballé, V.~Laparra, and E.~P. Simoncelli, ``Density {Modeling} of {Images}
  using a {Generalized} {Normalization} {Transformation},'' February 2016,
  arXiv:1511.06281 [cs]. [Online]. Available:
  \url{http://arxiv.org/abs/1511.06281}
\BIBentrySTDinterwordspacing

\bibitem{russakovsky_imagenet_2015}
\BIBentryALTinterwordspacing
O.~Russakovsky, J.~Deng, H.~Su, J.~Krause, S.~Satheesh, S.~Ma, Z.~Huang,
  A.~Karpathy, A.~Khosla, M.~Bernstein, A.~C. Berg, and L.~Fei-Fei,
  ``\BIBforeignlanguage{en}{{ImageNet} {Large} {Scale} {Visual} {Recognition}
  {Challenge}},'' \emph{\BIBforeignlanguage{en}{International Journal of
  Computer Vision}}, vol. 115, no.~3, pp. 211--252, December 2015. [Online].
  Available: \url{https://doi.org/10.1007/s11263-015-0816-y}
\BIBentrySTDinterwordspacing

\bibitem{lim_enhanced_2017}
\BIBentryALTinterwordspacing
B.~Lim, S.~Son, H.~Kim, S.~Nah, and K.~M. Lee, ``Enhanced {Deep} {Residual}
  {Networks} for {Single} {Image} {Super}-{Resolution},'' July 2017,
  arXiv:1707.02921 [cs]. [Online]. Available:
  \url{http://arxiv.org/abs/1707.02921}
\BIBentrySTDinterwordspacing

\bibitem{charbonnier_two_1994}
P.~Charbonnier, L.~Blanc-Feraud, G.~Aubert, and M.~Barlaud, ``Two deterministic
  half-quadratic regularization algorithms for computed imaging,'' in
  \emph{Proceedings of 1st {International} {Conference} on {Image}
  {Processing}}, vol.~2, November 1994, pp. 168--172 vol.2.

\bibitem{lai_deep_2017}
W.-S. Lai, J.-B. Huang, N.~Ahuja, and M.-H. Yang, ``Deep {Laplacian} {Pyramid}
  {Networks} for {Fast} and {Accurate} {Super}-{Resolution},'' in \emph{2017
  {IEEE} {Conference} on {Computer} {Vision} and {Pattern} {Recognition}
  ({CVPR})}, July 2017, pp. 5835--5843, iSSN: 1063-6919.

\bibitem{chen_symbolic_2023}
\BIBentryALTinterwordspacing
X.~Chen, C.~Liang, D.~Huang, E.~Real, K.~Wang, Y.~Liu, H.~Pham, X.~Dong,
  T.~Luong, C.-J. Hsieh, Y.~Lu, and Q.~V. Le, ``Symbolic {Discovery} of
  {Optimization} {Algorithms},'' February 2023, arXiv:2302.06675 [cs] version:
  2. [Online]. Available: \url{http://arxiv.org/abs/2302.06675}
\BIBentrySTDinterwordspacing

\bibitem{kingma2014adam}
D.~P. Kingma and J.~Ba, ``Adam: A method for stochastic optimization,''
  \emph{arXiv preprint arXiv:1412.6980}, 2014.

\bibitem{goblick_theoretical_1965}
T.~Goblick, ``Theoretical limitations on the transmission of data from analog
  sources,'' \emph{IEEE Transactions on Information Theory}, vol.~11, no.~4,
  pp. 558--567, 1965.

\bibitem{hu_model_2021}
X.~Hu, L.~Chu, J.~Pei, W.~Liu, and J.~Bian, ``Model complexity of deep
  learning: A survey,'' \emph{Knowledge and Information Systems}, vol.~63, pp.
  2585--2619, 2021.

\bibitem{weiHRSIDHighResolutionSAR2020}
S.~Wei, X.~Zeng, Q.~Qu, M.~Wang, H.~Su, and J.~Shi, ``{{HRSID}}: {{A
  High-Resolution SAR Images Dataset}} for {{Ship Detection}} and {{Instance
  Segmentation}},'' \emph{IEEE Access}, vol.~8, pp. 120\,234--120\,254, 2020.

\bibitem{UnsplashDatasetWorld2023}
``Unsplash {{Dataset}} | {{The}} world's largest open library dataset,''
  https://unsplash.com/data, March 2023.

\bibitem{wang_chestx-ray8_nodate}
X.~Wang, Y.~Peng, L.~Lu, Z.~Lu, M.~Bagheri, and R.~M. Summers, ``Chestx-ray8:
  Hospital-scale chest x-ray database and benchmarks on weakly-supervised
  classification and localization of common thorax diseases,'' in
  \emph{Proceedings of the IEEE conference on computer vision and pattern
  recognition}, 2017, pp. 2097--2106.

\end{thebibliography}

\vfill

\end{document}